\documentclass[iop]{emulateapj}

\usepackage{graphicx, amsmath, amssymb,natbib,multirow,epstopdf, bm}
\slugcomment{Submitted to the Astrophysical Journal}

\begin{document}
	
	\title{Hydrogen Burning on Accreting White Dwarfs: Stability, Recurrent
Novae, and the Post-Novae Supersoft Phase}
	\shorttitle{Hydrogen Burning on WDs}
	\author{William M. Wolf\altaffilmark{1}, Lars Bildsten\altaffilmark{1,2},
Jared Brooks\altaffilmark{1}, and Bill Paxton\altaffilmark{2}}
	\altaffiltext{1}{Department of Physics, University of California, Santa
Barbara, CA 93106}
	\altaffiltext{2}{Kavli Institute for Theoretical Physics, Santa Barbara, CA
93106}
	\shortauthors{Wolf et al.}
\begin{abstract}
	We examine the properties of white dwarfs (WDs) accreting hydrogen-rich
matter in and near the stable burning regime of accretion rates as modeled by
time-dependent calculations done with Modules for Experiments in Stellar
Astrophysics (\texttt{MESA}). We report the stability boundary for WDs of
masses between $0.51\ M_\odot$ and $1.34\ M_\odot$ as found via time-dependent
calculations. We also examine recurrent novae that are accreting at rates close
to, but below, the stable burning limit and report their recurrence times. Our
dense grid in accretion rates finds the expected minimum possible recurrence
times as a function of the WD mass. This enables inferences to be made about
the minimum WD mass possible to reach a specific recurrence time. We compare
our computational models of post-outburst novae to the stably burning WDs and
explicitly calculate the duration and effective temperature
($T_{\mathrm{eff}}$) of the post-novae WD in the supersoft phase. We agree with
the measured turnoff time - $T_{\mathrm{eff}}$ relation in M31 by Henze and
collaborators, infer WD masses in the 1.0-1.3 $M_\odot$ range, and predict
ejection masses consistent with those observed. We close by commenting on the
importance of the hot helium layer generated by stable or unstable hydrogen
burning for the short- and long-term evolution of accreting white dwarfs.
\end{abstract}	
	
\section{Introduction} 
\label{sec:introduction}
The outcome of accretion of hydrogen-rich material onto the surface of a white
dwarf (WD) is relevant to classical novae \citep{1978ARA&A..16..171G},
recurrent novae, supersoft sources (SSS) \citep{1992A&A...262...97V,
2007ApJ...663.1269N}, and even the single degenerate scenario (SDS) for type Ia
supernovae progenitors \citep{1982ApJ...253..798N, 1984ApJ...286..644N,
1998ApJ...496..376C}. The outcome depends on the mass of the accreting WD,
$M_{\mathrm{WD}}$, the accretion rate, \smash{$\dot{M}$}, and the core
temperature, $T_c$ \citep{1980A&A....85..295S, 2004ApJ...600..390T,
2005ApJ...623..398Y, 2007ApJ...663.1269N, 2007ApJ...660.1444S}. If the
accretion rate is too large, the burning can't match it, causing the rapidly
accreting matter to pile up into a red giant-like structure
\citep{1978ApJ...222..604P, 1979PASJ...31..287N}. At lower accretion rates,
hydrogen can be stably burned to helium at the same rate that is being accreted
\citep{1978ApJ...222..604P, 1980A&A....85..295S, 1982ApJ...259..244I,
1982ApJ...257..767F, 1983ApJ...264..282P, 1989ApJ...341..299L,
1998ApJ...496..376C, 2007ApJ...660.1444S, 2007ApJ...663.1269N}. If the
accretion rate is lower yet, the hydrogen supply rate is too low to match the
stably burning luminosity, so a low-luminosity accreting state is realized
while hydrogen accumulates until a thermonuclear runaway occurs, quickly
burning the hydrogen and driving a radius increase and mass loss from the WD
that appears as a classical or recurrent nova.

Understanding these phenomena first requires understanding the physics of
stable burning. Previous studies \citep{1980A&A....85..295S,
2007ApJ...663.1269N, 2007ApJ...660.1444S} assumed a steady burning state and
studied the stability of their solutions in response to linear perturbations.
While numerous time-dependent simulations of WDs accreting hydrogen-rich
material are already in the literature, many either started with the matter
pre-accreted and studied the ensuing outburst or selected initial conditions
which are not erased until several flashes have established asymptotic behavior
\citep{1982ApJ...259..244I}. Only \cite{1978ApJ...222..604P},
\cite{1979ApJ...230..832S}, \cite{1982ApJ...259..244I},
\cite{1989ApJ...341..299L}, \cite{1993ApJ...406..220S},
\cite{1994ApJ...424..319K}, \cite{1998ApJ...496..376C}, and
\cite{2005ApJ...623..398Y} examined the time-dependent problem for durations
long enough to observe multiple flashes or stable burning. To date, there is no
comprehensive time-dependent study of WDs accreting solar composition material
over the full range of the stable burning regime for a large range of WD
masses. \cite{2005ApJ...623..398Y} was the most complete prior effort, but did
not calculate a dense grid in \smash{$\dot{M}$} space near the lower stability
boundary.

WDs accreting just below the stability boundary will go through periodic
hydrogen shell flashes on relatively short timescales, or recurrent novae
(RNe). There are currently ten known RNe in our galaxy
\citep{2010ApJS..187..275S} with recurrence times on the order of decades.
Those RNe with shortest recurrence times have been understood to be massive WDs
(see Figure 9 of \citealt{1982ApJ...253..798N}). Since the measured time
between outbursts is often an important factor in estimating
\smash{$M_{\mathrm{WD}}$}, we require a solid understanding of the
mass-recurrence time relation near the lower stability boundary.

At lower \smash{$\dot{M}$}'s ($\lesssim 10^{-8}\ M_\odot\,\mathrm{yr}^{-1}$),
WDs undergo classical nova cycles whose recurrence times are too long to
measure on human timescales \citep{2005ApJ...623..398Y}. Mass determinations of
such systems must then rely on other observed parameters. After a CN outburst,
the ejected mass is optically thick, obscuring the view of the hot WD below.
After a turn-on time, $t_{\mathrm{on}}$, the ejecta becomes optically thin,
revealing an SSS. Still later, at some turn-off time $t_{\mathrm{off}}$ after
the outburst, the X-ray luminosity powered by burning in the hydrogen-rich
remnant \citep{1974ApJS...28..247S} decreases and the nova event is over.
\cite{2010ApJ...709..680H} have offered a way to fit the observed timescales to
models in order to infer $M_{\mathrm{WD}}$. \cite{1998ApJ...503..381T} and
\cite{2005A&A...439.1061S} argue that the mass left in the hydrogen-rich WD
envelope after mass loss is determined primarily by $M_{\mathrm{WD}}$ and
secondarily by the composition of the envelope. This remnant envelope mass is
expected to undergo stable hydrogen burning in the post-outburst SSS phase at
nearly constant luminosity. If this remaining envelope mass and luminosity are
known as a function $M_{\mathrm{WD}}$, the duration of the SSS phase can be
predicted, allowing a correlation between the measured turn-off time of a CN
and $M_{\mathrm{WD}}$. With the increasingly large samples of CNe like that of
\cite{Henze:2011bn} in M31 and \cite{2011ApJS..197...31S} in our own galaxy, we
can now test these methods on a meaningful number of CNe.

In this paper, we present models of WDs with masses ranging from
$M_{\mathrm{WD}}=0.51\ M_\odot$ to 1.34 $M_\odot$ accreting solar composition
material as simulated by \texttt{MESA} \citep{2011ApJS..192....3P,
2013ApJ...762....8D,2013ApJS..208....4P}. The conditions for stable hydrogen burning
are found, as are the characteristics of unstable models. We compare recurrence
times to the previous results from \cite{1982ApJ...259..244I},
\cite{1989ApJ...341..299L}, \cite{1998ApJ...496..376C}, and
\cite{2005ApJ...623..398Y}. We start in \S
\ref{sec:simulation_details_and_model_building} by discussing the input physics
used to produce the accreting models. Then in \S
\ref{sec:steadily_burning_models}, we present the relevant background on stable
burning as well as the characteristics of our steadily (and stably) burning
models. We investigate unstable burning on WDs accreting at rates near, but
below, the stable boundary in \S \ref{sec:unstable_burning} with comparisons to
previous time-dependent calculations explored in \S
\ref{sec:comparisons_to_other_studies}. We study applications to classical
novae in \S \ref{sec:post_outburst_novae}, and further implications and
questions are addressed in \S \ref{sec:concluding_remarks}, where we comment on
the inevitable flashes in the accumulating helium layer.


\section{Simulation Details and Model Building } 
\label{sec:simulation_details_and_model_building}
	Initial WD models were created in \texttt{MESA} by evolving stars between 4
and 12 $M_\odot$ from ZAMS through the main sequence, RGB, and AGB through to
the white dwarf cooling track. Typically when this is done, very small time
steps are required to get through the thermal pulses during the AGB phase. To
get around this, the convection in the outer envelope is artificially made more
efficient so that the full computations need not be followed
\citep{2013ApJS..208....4P}. Additionally, the larger initial models used
enhanced winds to speed up the process, which is why they didn't undergo core
collapse. These processes do not impact our results since the physics of
interest is in the accreted envelope and nearly independent of the degenerate
interior. This process was also used and discussed in
\cite{2013ApJ...762....8D}.
	
	After the initial WD models were created, they were cooled to a central
temperature of $T_c=3\times 10^7\ \mathrm{K}$, hot enough so that the initial
flash is not too violent. The results of \cite{2005ApJ...623..398Y} show that
for accretion rates in the stable regime, the accumulated Helium layer is at
$T\approx 10^8\ \mathrm{K}$, making the results of hydrogen-rich accretion
nearly independent of $T_c<10^8\ \mathrm{K}$, a result we also justify in \S
\ref{sec:steadily_burning_models}. We assume that all WDs are not rotating and
that convective overshoot does not occur. To model the nuclear burning, we used
\texttt{MESA}'s \verb!cno_extras_o18_to_mg26_plus_fe56! network, which accounts
for hot CNO burning as well as other heavier isotopes' presence. Radiative
opacities are from the OPAL tables \citep{1993ApJ...412..752I,
1996ApJ...464..943I}. The spatial and temporal resolutions were adjusted to
finer and finer levels until no more substantive changes were observed in the
stability/instability boundary or in the reported observables (recurrence
times, burning layer temperatures, envelope masses, etc.). This typically
resulted in models with between 7000 and 10000 mass zones that are dynamically
sized in space and time so that a burning region in an active nova or steady
burner is well-resolved, typically occupying around half of the mass zones.

	The accreted material has solar composition, with $X=0.70$, $Y=0.28$, and
metal fractions taken from \cite{2003ApJ...591.1220L}, though the OPAL
opacities assume a different set of metal fractions for solar composition.
Initializing the accretion often required irradiating the atmosphere before
starting accretion so as to ease the thermal readjustment of the outer layers.
Any unphysical effects this would have on the model are undone after the
ensuing flash(es) that erase the initial conditions
\citep{1978ApJ...222..604P,1979ApJ...230..832S, 1998ApJ...496..376C}. The first
flash heats the outer layers so that the irradiation is no longer needed for
computational convenience. After several flashes (in unstable models) or
hydrogen sweeping times, \smash{$\Delta M_{\mathrm{H}}/(X\dot{M})$}, where
$\Delta M_{\mathrm{H}}$ is the total hydrogen mass (for stably burning models),
any hydrogen present in the envelope during the initial accumulation phase has
already been burned to helium or ejected. It is after this initial ``memory
erasing'', with irradiation deactivated that we begin our exploration.

For this study, we employ two mass loss prescriptions: super Eddington winds
and Roche lobe overflow. For the purposes of our calculation, the RNe systems
are assumed to be wide binaries. This is not the case for all RNe. With this
assumption, the only active mass loss prescription is the super Eddington wind
scheme described in \cite{2013ApJ...762....8D}. In this prescription,
winds are only active if the photosphereic luminosity of the star exceeds an
effective Eddington luminosity which is a mass-average of the local Eddington
luminosity from the outer-most cell down to where the optical depth first
exceeds 100. The excess luminosity over this effective Eddington luminosity
comes in the form of mass ejection moving at the surface escape velocity. Both
of these mass ejection scenarios take place over an extended period of time
until $L<L_{\mathrm{Edd}}$ or $R<R_{\mathrm{RL}}$, usually indicating the end
of the nova's excursion to the red in the HR diagram. The tighter binaries in
which CNe are found can result in either of these ejection scenarios, so we
present results with both assumptions in \S \ref{sec:post_outburst_novae},
where we also describe the Roche lobe wind prescription in more detail. The
inlists for these simulations are available on \texttt{http://www.mesastar.org}.
	
\begin{figure}[htb]
		\centering
		\includegraphics[width=\columnwidth]{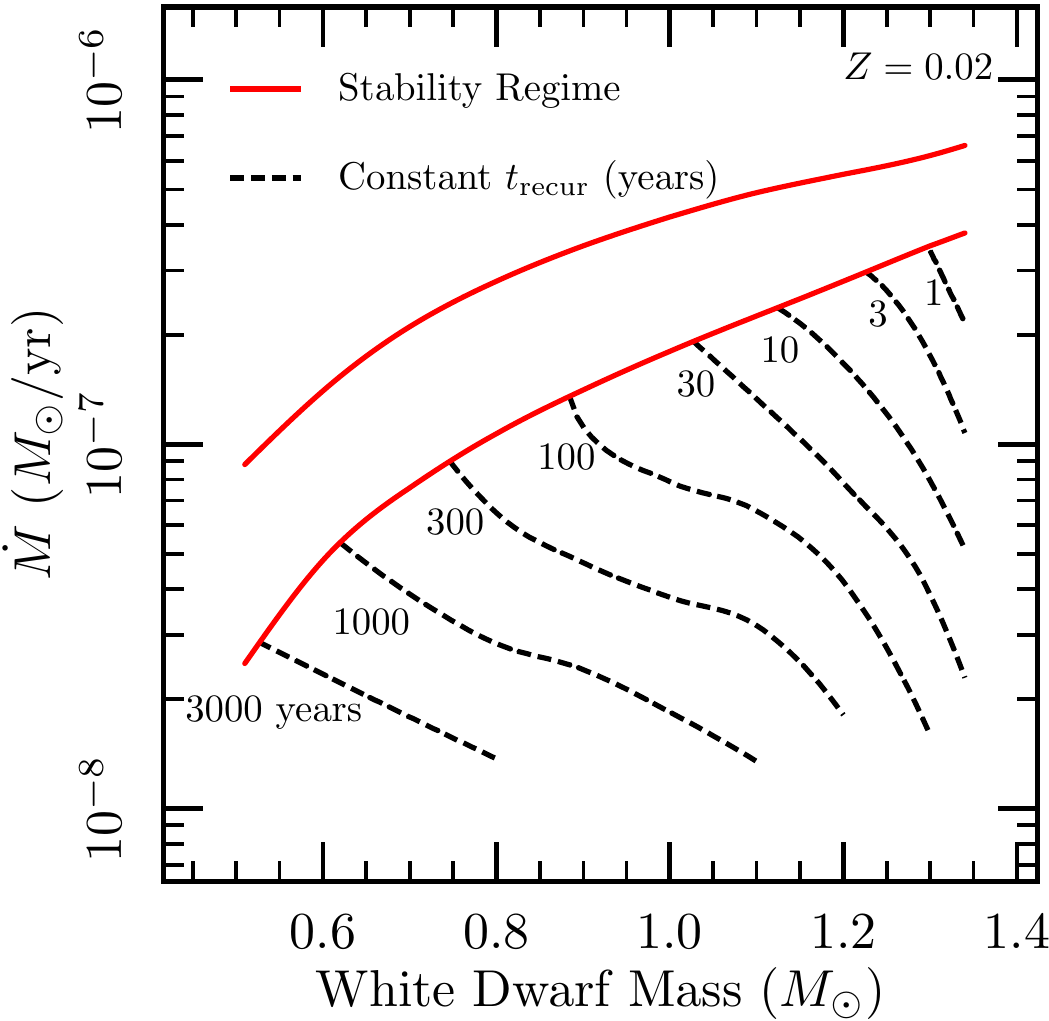}
		\caption{Location of the stable-burning regime on the
		\smash{$M-\dot{M}$} plane. The lower red line represents
		\smash{$\dot{M}_{\mathrm{stable}}$}, the lowest possible accretion rate
		which exhibits stable and steady burning for a given WD mass. The upper
		red line gives the highest such accretion rate,
		\smash{$\dot{M}_{\mathrm{RG}}$}. WDs accreting at rates above
		\smash{$\dot{M}_{\mathrm{RG}}$} will still burn stably, but not at the
		rate that the matter is being accreted, causing the matter to pile up,
		forming a red giant-like structure. Also shown in dashed black
		lines are lines of constant recurrence time (in years) for recurrent
		novae as interpolated from our grid. }
		\label{fig:2}
	\end{figure}

\begin{figure}[htb]
		\centering 
		\includegraphics[width=\columnwidth]{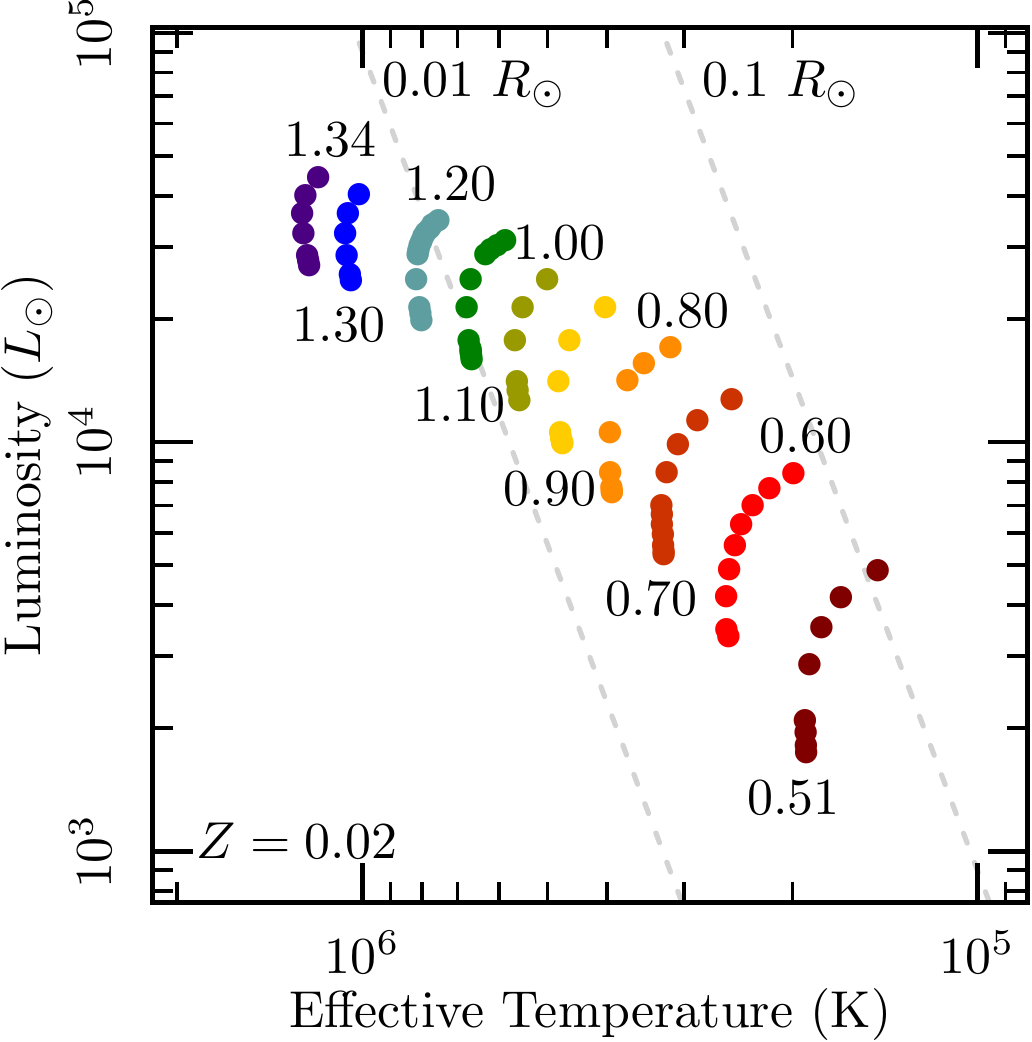}
		\caption{Positions of stable burning models of accreting WDs on the HR
		diagram. The different colors indicate the different WD masses used in the
		simulations. Lines of constant radii are drawn for $R=0.01\ R_\odot$ and
		$0.1\ R_\odot$. The most luminous point for each mass corresponds to the WD
		accreting at \smash{$\dot{M} = \dot{M}_{\mathrm{stable}} +
		0.8(\dot{M}_{\mathrm{RG}}-\dot{M}_{\mathrm{stable}})$}. At accretion rates
		close to \smash{$\dot{M}_{\mathrm{RG}}$}, the radius becomes ill-defined as
		the envelope slowly expands, so we only report those WDs with
		well-established radii.}
		\label{fig:1}
	\end{figure}

	We explore a wide range of WD masses for accretion rates in and near the
stable burning regime, so we prepared C-O WDs of masses 0.51 $M_\odot$, $0.60\
M_\odot$, $0.70\ M_\odot$, $0.80\ M_\odot$, $1.00\ M_\odot$, and O-Ne WDs of
masses $1.10\ M_\odot$, $1.20\ M_\odot$, $1.30\ M_\odot$, and $1.34\ M_\odot$.
Then, using earlier studies of stability regimes \citep{1980A&A....85..295S,
2007ApJ...663.1269N, 2007ApJ...660.1444S}, we chose accretion rates within and
near the stable burning regime. For each WD we then relaxed the accretion rate
to the desired rate and allowed the model to evolve for at least 30 envelope
turnover times to erase all history, typically accreting $<10^{-3}\ M_\odot$.
While the study of much longer-term accretion is certainly warranted
\citep{1998ApJ...496..376C, 1999ApJ...521L..59P}, we wanted to initially avoid
introducing significant temporal changes to the models resulting from their
increasing mass due to secular accumulation of helium. At the end of this
accretion period, we take the measurements shown in Figures~\ref{fig:2} and
\ref{fig:1}.
	
\section{Steadily Burning Models } 
\label{sec:steadily_burning_models}
	The conditions for stable burning of hydrogen-rich material are detailed in
\cite{2007ApJ...660.1444S} for a simple one-zone model of burning. The
qualitative results are that steady-state burning becomes stable when an
increase in temperature causes the cooling rate to increase more than the
energy generation rate. At accretion rates below a certain critical value,
\smash{$\dot{M}_{\mathrm{stable}}$}, a temperature perturbation would cause the
nuclear heating rate to grow faster than the cooling rate causing a
thermonuclear runaway that burns the fuel at a rate faster than accretion,
triggering a limit-cycle of accumulation and explosion (i.e.~novae).
Time-dependent calculations naturally reveal the lower stability bound, as
unstable periods of hydrogen burning manifest themselves. Table~\ref{tab:1}
summarizes these results, indicating the lower limiting accretion rate for
stable burning, \smash{$\dot{M}_{\mathrm{stable}}$}, the total mass of
hydrogen, $\Delta M_{\mathrm{H}}$, the hydrogen sweeping time,
\smash{$t_{\mathrm{sweep}}=\Delta M_{\mathrm{H}} /
(X\dot{M}_{\mathrm{stable}})$}, and the luminosity. Additionally, we show the
pressure, density, temperature, fraction of pressure due to gas pressure, and
hydrogen mass fraction at the point where the exiting luminosity is half of the
total luminosity. Finally, we show the thickness of the shell from the
half-luminosity point to the surface of the WD as a fraction of the total
radius, $R$, demonstrating that these burning shells are only marginally thin,
enhancing their stability. For the range of masses shown in Figure~\ref{fig:2},
the lower red line represents the values of
\smash{$\dot{M}_{\mathrm{stable}}$}. WDs in the region below the lower red line
in Figure~\ref{fig:2} would be recurrent novae.
	
	The upper edge of the stable regime is more subtle.
\cite{1982ApJ...257..767F} and \cite{1982ApJ...259..244I} note that for any WD,
there is a maximum envelope mass that can sustain steady-state burning. This
corresponds with a plateau in luminosity that is related to the core
mass-luminosity relation first found by \cite{1970AcA....20...47P} in AGB
cores. In transitioning to a more AGB-like envelope, the luminosity is limited
to a maximum value governed by the core mass, and so increasing
\smash{$\dot{M}$} just causes more matter to pile on to the envelope while
steady-state burning at the plateau luminosity continues at the base of the
envelope. From this plateau luminosity, we then identify a hydrostatic upper
limit to the accretion rate, \smash{$\dot{M}_{\mathrm{RG}} =
L_{\mathrm{plateau}} / XQ_{\mathrm{CNO}}$}. \cite{2007ApJ...660.1444S}
explicitly showed that this leads to an upper bound on the stable regime that
is tightly constrained to \smash{$\dot{M}_{\mathrm{RG}}\approx
3\dot{M}_{\mathrm{stable}}$}. It is always the case that
\smash{$\dot{M}_{\mathrm{RG}}$} is a stronger upper limit on the accretion rate
than that set by the Eddington luminosity, \smash{$\dot{M}_{\mathrm{Edd}} =
L_{\mathrm{Edd}} / XQ_{\mathrm{CNO}}$}, but in our code it often manifested
itself by triggering super Eddington winds since the increase in radius caused
the opacity in the outer layers to diverge from pure electron scattering.
	
	 For a given mass, increasing \smash{$\dot{M}$ to $\dot{M}_{\mathrm{RG}}$}
	 causes the WD to travel along a path in the HR diagram to higher $L$ and
	 $T_{\mathrm{eff}}$ until it hits a ``knee'', at which point the luminosity
	 continues to grow, but the effective temperature decreases, indicating a
	 radial expansion of the envelope. This knee can be seen in
	 Figure~\ref{fig:1}. The regime inhabited by these stably and steadily burning
	 WDs in the HR diagram is also known to hold many of the supersoft sources
	 \citep{2007ApJ...663.1269N}, making these stable burners excellent candidates
	 as the source of the soft X-rays. At high enough \smash{$\dot{M}$}'s,
	 hydrogen burning cannot burn at the same rate as accretion, causing a radial
	 expansion in the envelope and a build-up of hydrogen
	 \citep{1979PASJ...31..287N}. The upper line in Figure~\ref{fig:2} represents
	 models with the highest \smash{$\dot{M}$} that exhibit steady-state burning
	 of hydrogen at the accreted rate, \smash{$\dot{M}_{\mathrm{RG}}$}. WDs in the
	 region above the upper line in Figure~\ref{fig:2} will still burn hydrogen at
	 a constant rate (albeit more slowly than it is being accreted), and their
	 envelopes will grow until optically-thick winds or Roche-lobe overflow can
	 slow the accretion rate \citep{1996ApJ...470L..97H}. We don't investigate
	 these systems in our study except to find the value of
	 \smash{$\dot{M}_{\mathrm{RG}}$} for each mass.

	\begin{deluxetable*}{ccccccccccc}
		\tablecolumns{11}
		\tablewidth{0pc}
		\tablecaption{Properties of Stably Burning WDs at the Stability Boundary}
		\tablehead{
			\colhead{} & \multicolumn{4}{c}{Stability Boundary} &
				\multicolumn{6}{c}{At Half Total Luminosity}\\
			\cline{2-5} \cline{6-11}
			\colhead{$M_{\mathrm{WD}}$}  & \colhead{$\dot{M}_{\mathrm{stable}}$} &
				\colhead{$\Delta M_{\mathrm{H}}$} & \colhead{$t_{\mathrm{sweep}}$} &
				\colhead{$L$} & \colhead{$P$} & \colhead{$\rho$} & \colhead{$T$} &
				\colhead{$\beta$} & \colhead{$X_{\mathrm{H}}$} &
				\colhead{$\Delta R/R$}\\
			\colhead{($M_\odot$)} & \colhead{($10^{-7}M_\odot\,\mathrm{yr}^{-1}$)} &
				\colhead{($M_\odot$)} & \colhead{(yr)} & \colhead{($10^3\,L_\odot$)} &
				\colhead{($10^{17}\,\mathrm{dyne/cm^2}$)} &
				\colhead{($\mathrm{g/cm^3}$)} & \colhead{($10^7\,\mathrm{K}$)} &
				- & - & -
		}
		\startdata
		0.51 & 0.25 & $4.9\times 10^{-5}$ & 2810 & 1.74 & 2.31 & 47.5 & 5.08 &
			 0.928 & 0.258 & 0.558\\
		0.60 & 0.48 & $2.4\times 10^{-5}$ & 720  & 3.36 & 2.25 & 36.0 & 5.69 &
			 0.882 & 0.335 & 0.496\\
		0.70 & 0.76 & $1.2\times 10^{-5}$ & 226  & 5.32 & 2.48 & 36.7 & 6.24 &
			 0.846 & 0.286 & 0.430\\
		0.80 & 1.07 & $6.7\times 10^{-6}$ & 88.7 & 7.51 & 2.74 & 35.9 & 6.72 &
			 0.812 & 0.290 & 0.376\\
		0.90 & 1.41 & $3.5\times 10^{-6}$ & 35.4 & 9.89 & 3.15 & 37.0 & 7.22 &
			 0.783 & 0.291 & 0.323\\
		1.00 & 1.80 & $2.0\times 10^{-6}$ & 15.5 & 12.7 & 3.63 & 38.9 & 7.72 &
			 0.753 & 0.280 & 0.285\\
		1.10 & 2.40 & $9.6\times 10^{-7}$ & 6.08 & 15.9 & 4.37 & 42.1 & 8.32 &
			 0.723 & 0.272 & 0.244\\
		1.20 & 2.80 & $4.4\times 10^{-7}$ & 2.24 & 19.8 & 5.41 & 44.8 & 9.05 &
			 0.687 & 0.284 & 0.210\\
		1.30 & 3.50 & $1.3\times 10^{-7}$ & 0.534 & 24.9 & 7.66 & 52.4 & 10.2 &
			 0.644 & 0.293 & 0.168\\
		1.34 & 3.80 & $6.0\times 10^{-8}$ & 0.226 & 27.0 & 9.76 & 61.0 & 10.9 &
			 0.630 & 0.290 & 0.141
		\enddata
		\label{tab:1}
	\end{deluxetable*}

\begin{figure} 
	\centering
	\includegraphics[width=\columnwidth]{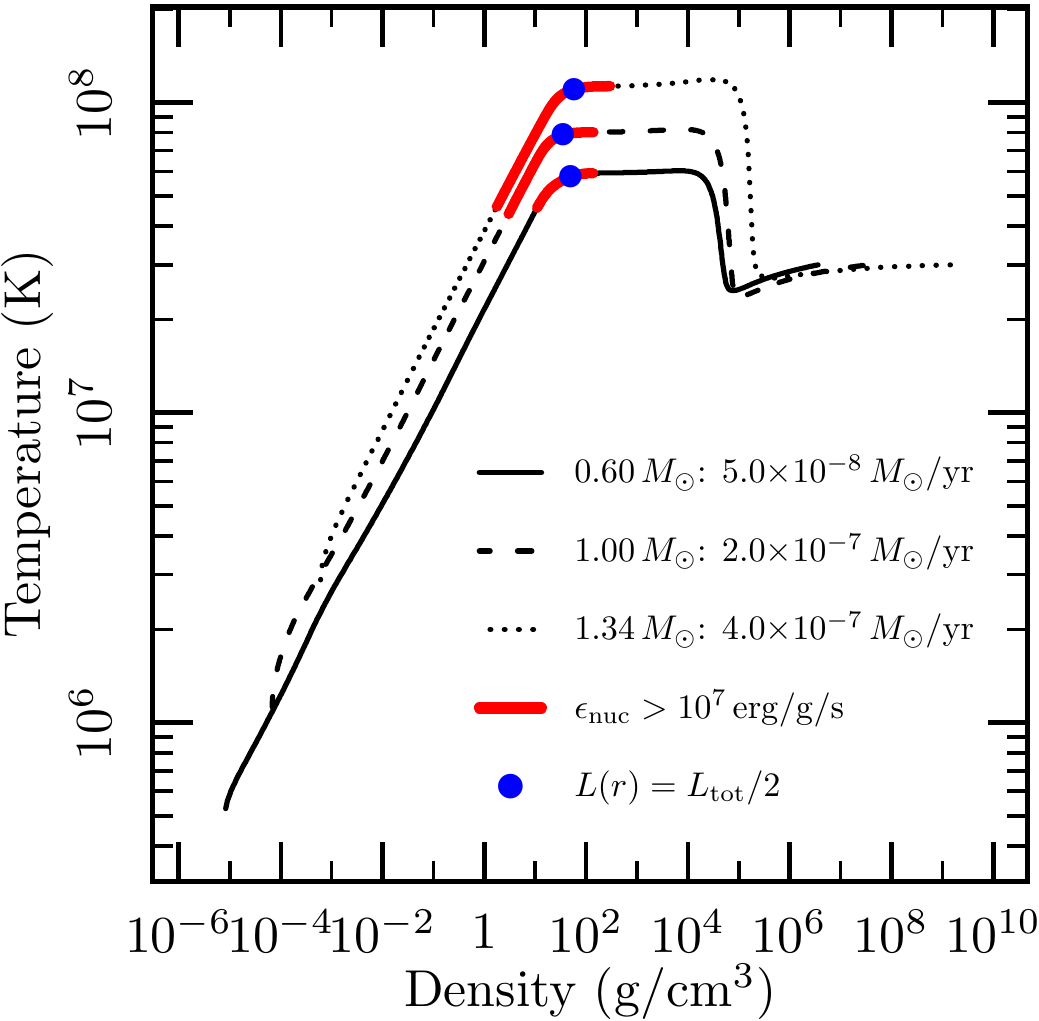}
	\caption{The temperature-density profile of $0.60\ M_\odot$, $1.00\ M_\odot$,
and $1.34\ M_\odot$ WDs accreting at \smash{$\dot{M}= 5.0\times 10^{-8}\
M_\odot\,\mathrm{yr}^{-1}$}, \smash{$\dot{M}=2.0\times 10^{-7}\
M_\odot\,\mathrm{yr}^{-1}$}, and \smash{$\dot{M}=4.0\times 10^{-7}\
M_\odot\,\mathrm{yr}^{-1}$}, respectively. Areas of significant
hydrogen burning are marked, as well as the point where the exiting
luminosity is half of the total luminosity of the star.}
	\label{fig:trho}
\end{figure}

	The internal structures of $0.60\ M_\odot$, $1.00\ M_\odot$, and $1.34\
M_\odot$ WDs accreting at \smash{$\dot{M}=5.0\times 10^{-8}\ M_\odot\,\mathrm{yr}^{-1}$},
$2.0\times 10^{-7}\ M_\odot\,\mathrm{yr}^{-1}$, and $4.0\times 10^{-7}\
M_\odot\,\mathrm{yr}^{-1}$, respectively, are shown in Figure~\ref{fig:trho}.

\begin{figure}
	\centering
	\includegraphics[width=\columnwidth]{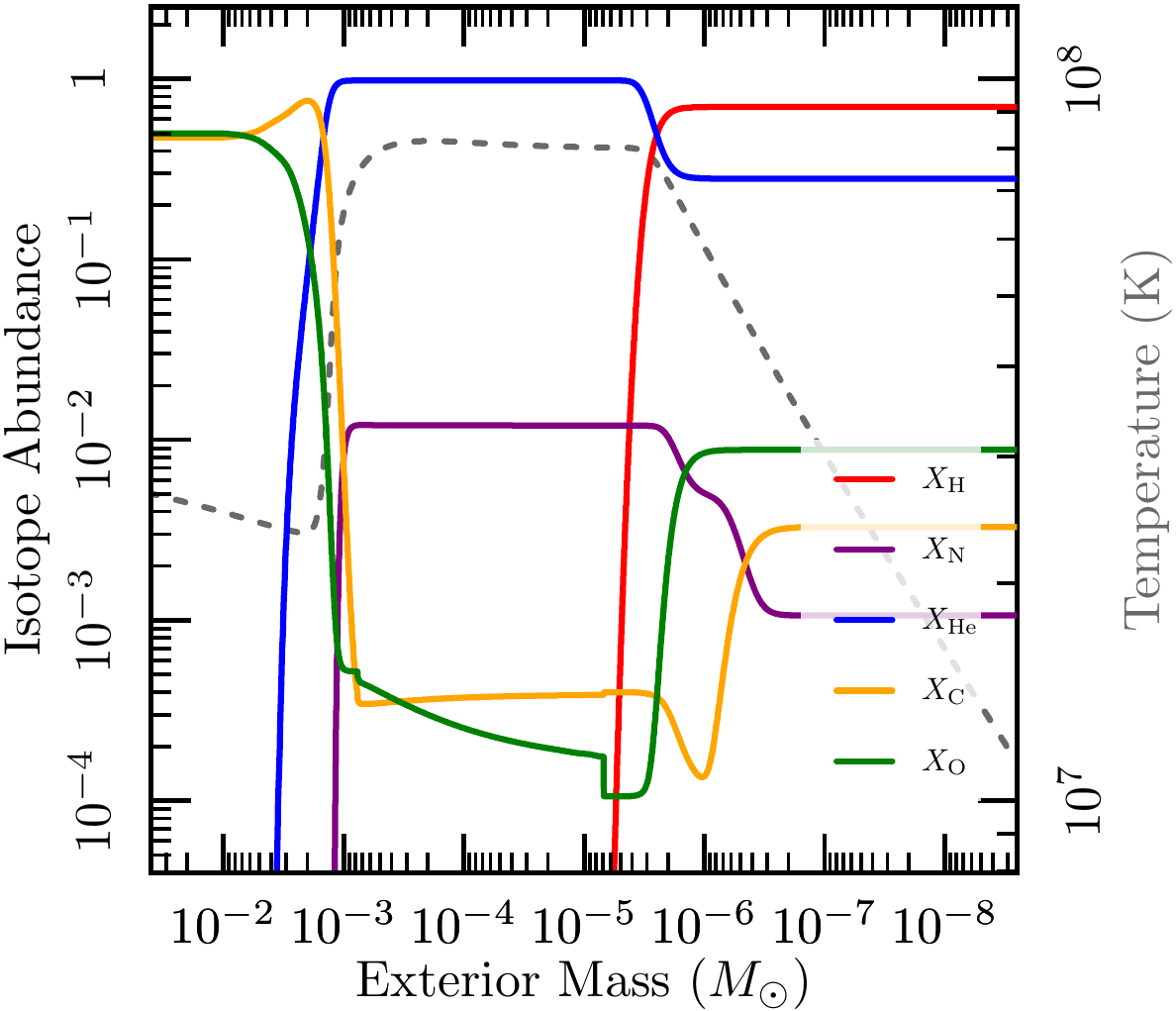}
	\caption{The abundance profile of a $1.00\ M_\odot$ WD accreting at
\smash{$\dot{M}=2.0\times 10^{-7}\ M_\odot\,\mathrm{yr}^{-1}$}. The temperature
profile (dashed line) has also been included to show that the hottest region is
the layer of helium ash that dominates the burning conditions rather than the
cooler core.}
	\label{fig:abundances}
\end{figure}

	The cores are largely isothermal and degenerate, but on top of them is a hot
helium layer that is the ash of the stable burning. Above the ash is the
radiative hydrogen envelope with most of the burning occurring just above the
Helium ash. To further illustrate the presence of a thick, hot helium
layer, Figure~\ref{fig:abundances} shows the elemental abundance and
temperature profile of the same $1.00\ M_\odot$ WD shown in
Figure~\ref{fig:trho}. Additionally, we see the expected pattern of a
hydrogen-helium transition zone above the hot ash, coinciding with a rising
${}^{14}$N mass fraction due to CNO burning.

	From Figures~\ref{fig:trho} and \ref{fig:abundances} as well as
Table~\ref{tab:1}, we see that there is a temperature $T_{\mathrm{stable}}$ at
which the stable burning occurs, and that $T_{\mathrm{stable}}$ is an
increasing function of both \smash{$\dot{M}$} and $M_{\mathrm{WD}}$. We can
explain this dependence with a few assumptions about the nature of the
hydrogen-rich envelope. We expect the burning to occur at a depth where the
burning timescale, $t_{\mathrm{burn}}\sim
Q_{\mathrm{CNO}}/\epsilon_{\mathrm{CNO}}$ (where $\epsilon_{\mathrm{CNO}}$ is
the nuclear energy generation rate per unit mass and $Q_{\mathrm{CNO}}$ is the
amount of energy released per unit mass of hydrogen undergoing complete CNO
burning) is approximately equal to the accretion timescale,
\smash{$t_{\mathrm{acc}}=\Delta M/\dot{M}$}. If we assume a thin shell, where
the pressure is approximately $P= GM_{\mathrm{WD}}\Delta M/(4\pi
R_{\mathrm{core}}^4)$, (where $R_{\mathrm{core}}$ is the radius at the base of
the hydrogen-rich envelope) we get the burning condition to be
\smash{$\epsilon_{\mathrm{CNO}} = (Q_{\mathrm{CNO}}G\dot{M}M_{\mathrm{WD}})/(P
R_{\mathrm{core}}^4)$}. Furthermore, the opacity in the hydrogen-rich layer is
dominated by electron scattering and the envelope is radiative, so the
accretion rate can be related to temperature and pressure via
\smash{$XQ_{\mathrm{CNO}}\dot{M} = L \propto M_{\mathrm{WD}} T^4/P$}. For our
uses, we want to eliminate pressure, so we use \smash{$P\propto
M_{\mathrm{WD}}T^4/\dot{M}$}. Finally, we expand $\epsilon_{\mathrm{CNO}}$ as a
power law in temperature, and with the assumption pressure is due primarily to
gas pressure, we may use \smash{$\epsilon_{\mathrm{CNO}}\propto \rho T^\nu =
(P/T)T^\nu = M_{\mathrm{WD}}T^{\nu+3}/\dot{M}$}. Putting this all together we
find
	\begin{equation}
		\label{eq:burning_condition4} T_{\mathrm{stable}} \propto
			 \dot{M}^{3/(\nu+7)} M_{\mathrm{WD}}^{-1/(\nu+7)}
			 R_{\mathrm{core}}^{-4/(\nu + 7)},
	\end{equation}
	where $\nu = 23.58/T_7^{1/3} - 2/3$ and $T_7 = T/10^7\ \mathrm{K}$ as shown
in \cite{Hansen:2004vx}. Note that $R_{\mathrm{core}}$ is negatively correlated
with $M_\mathrm{WD}$ but positively (though weakly) correlated with
\smash{$\dot{M}$}. As a result, we expect $T_{\mathrm{stable}}$ to
\emph{increase} with increasing $M_{\mathrm{WD}}$ at constant \smash{$\dot{M}$}
through the implicit $R_{\mathrm{core}}$-dependence. For a fixed mass, the
radius is approximately constant with changing \smash{$\dot{M}$}, so
$T_{\mathrm{stable}}$ is only depending on a small power of \smash{$\dot{M}$}.
Using a prefactor of $3.5\times 10^8\ \mathrm{K}$ (assuming $M_{\mathrm{WD}}$
and $R_{\mathrm{core}}$ are measured in solar units and \smash{$\dot{M}$} in
$M_\odot\,\mathrm{yr}^{-1}$), Equation~\eqref{eq:burning_condition4} and the
$\nu-T_7$ relation yield temperatures at the point of peak burning accurately
to within 20\%. For intermediate masses (0.6\,$M_\odot \leq M_{\mathrm{WD}}
\leq 1.2\,M_\odot$), the calculated temperatures are typically well within 10\%
of the simulated values.


\section{Unstable Burning} 
\label{sec:unstable_burning}
	At accretion rates below the stable burning boundary
(\smash{$\dot{M}<\dot{M}_{\mathrm{stable}}$}) indicated in Table~\ref{tab:1},
the WDs undergo periodic hydrogen flashes. For accretion rates near the stable
boundary, these flashes lead to little mass loss from the system. Higher mass
WDs experience shorter recurrence times for a given \smash{$\dot{M}$}.
Equivalently, the ignition mass (the mass of accreted material at which a
runaway occurs) is smaller for larger core masses where the higher surface
gravity allows for higher pressures with less mass accumulation. If we look at
the ``first unstable model'' (the model with
\smash{$\dot{M}\lesssim\dot{M}_{\mathrm{stable}}$}), we can identify the
minimum recurrence time (or equivalent ignition mass) for that core mass.
Before exploring these boundary cases, we should justify our assumption that
such a limiting configuration exists

	\cite{1983ApJ...264..282P} examined flashes on hydrogen-accreting
compact objects with a simple one-zone model using linear stability analysis.
His analysis showed that as the accretion rate is decreased, the steady state
models go from stable (perturbations die exponentially) through quasi-stable
(perturbations act as damped oscillators), quasi-unstable (perturbations
oscillate with increasing amplitude), and finally fully unstable phases. When
simulating the unstable-to-stable flash transition, though, he found that the
transformation was very rapid. The one-zone models gave large amplitude flashes
(i.e.\ novae) until the accretion rate reached the stable accretion rate at
which point the model switched over to stable and steady burning with very
little weakening of the flashes. In other words, there is essentially a
discontinuity in the stability of the burning very near
\smash{$\dot{M}_{\mathrm{stable}}$}.

\begin{figure} 
		\centering
		\includegraphics[width=\columnwidth]{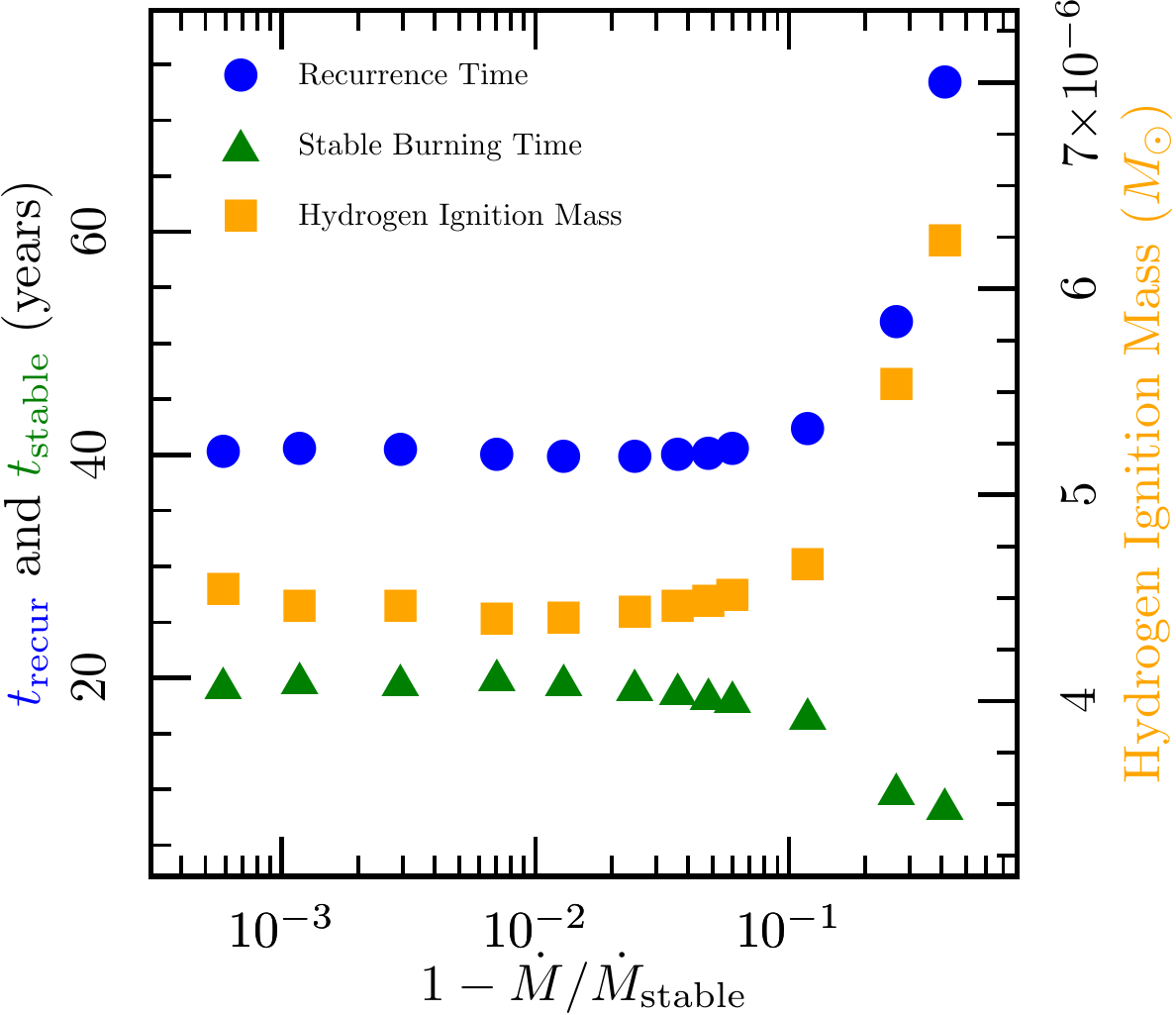}
		\caption{Convergence of the recurrence time, hydrogen ignition mass, and
		stable burning duration as \smash{$\dot{M}$} approaches
		\smash{$\dot{M}_{\mathrm{stable}}$} for $M_{\mathrm{WD}}=1.00\ M_\odot$.}
		\label{fig:first_unstable}
	\end{figure}

	In Figure~\ref{fig:first_unstable} we plot a high resolution grid of
\smash{$\dot{M}$}'s performed on the $1.00\ M_\odot$ model, demonstrating that
as \smash{$\dot{M}$} approaches \smash{$\dot{M}_{\mathrm{stable}}$}, the
igntion mass and recurrence times indeed approach nearly constant values of
$\Delta M_{\mathrm{H}} \approx 4.4\times 10^{-6}\ M_\odot$ and
$t_{\mathrm{recur}} \approx$ 40 days. While we didn't compute this fine of a
grid for each mass tested, we obtained the stable/unstable boundary resolved to
within five percent of \smash{$\dot{M}_{\mathrm{stable}}$}. This is precise
enough for identifying limiting recurrence times and hydrogen ignition masses.
	
	 An observer can then use an observed nova recurrence time to infer a
	 minimum core mass. Table~\ref{tab:2} lists the recurrence times for each
	 first unstable model. There we also list the peak temperature in the helium
	 layer in the low-luminosity state as well as the peak temperature in the
	 burning layer as the convective burning zone develops. The temperature in
	 the helium layer during the low-luminosity state is always close to the
	 extrapolated stable burning temperature (the temperature that would be
	 expected by the empirical power law fit at that \smash{$\dot{M}$}), though
	 the two temperatures do not track monotonically due to the varying helium
	 mass from model to model.

\begin{figure}
		\centering
		\includegraphics[width=\columnwidth]{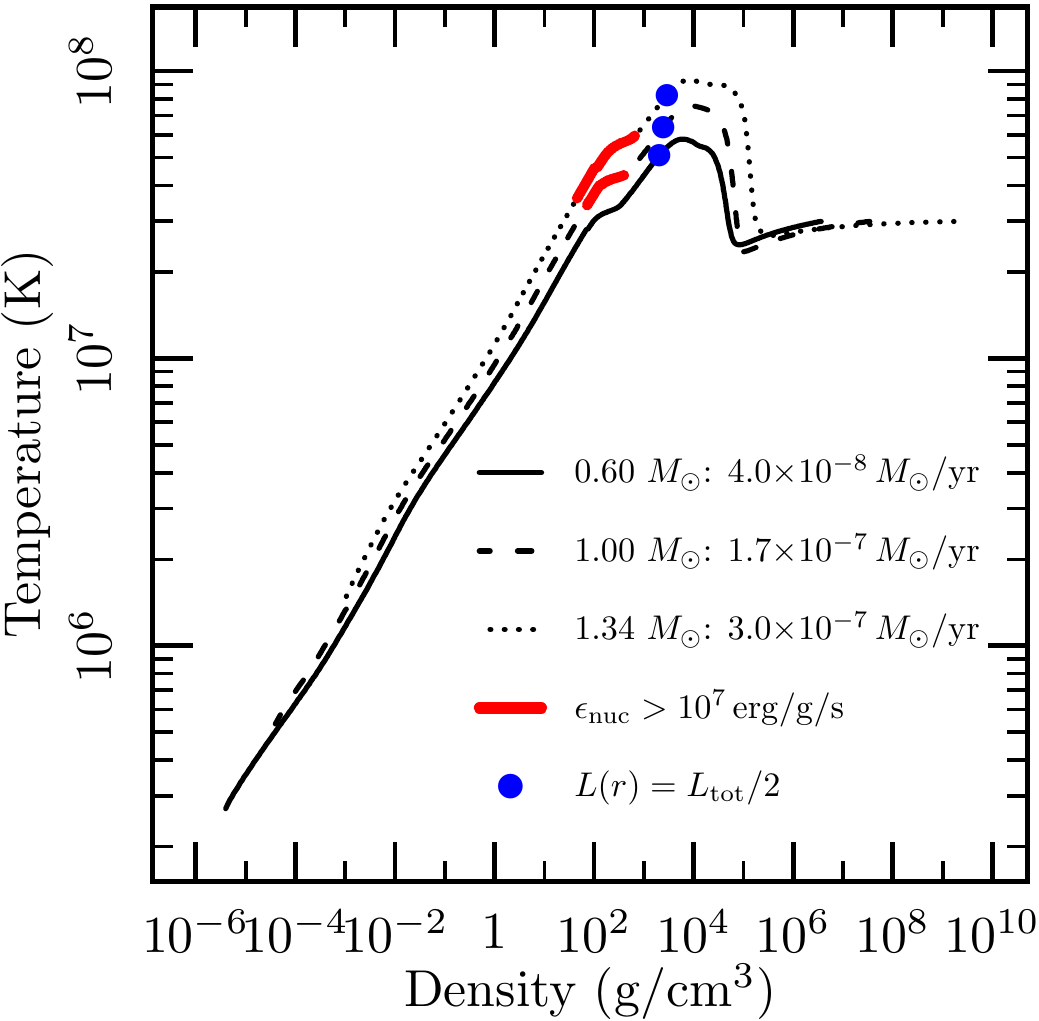}
		\caption{The temperature-density profile of $0.60\ M_\odot$, $1.00\
		M_\odot$, and $1.34\ M_\odot$ WDs accreting at \smash{$\dot{M}= 3.0\times
		10^{-8}\ M_\odot\,\mathrm{yr}^{-1}$}, \smash{$\dot{M}=1.7\times 10^{-7}\
		M_\odot\,\mathrm{yr}^{-1}$}, and \smash{$\dot{M}=3.0\times 10^{-7}\
		\,\mathrm{yr}^{-1}$}, respectively. Areas of significant CNO burning are
		marked, as well as the point where the exiting luminosity is half of the
		total luminosity of the star. These profiles correspond to the quiescent
		(pre-nova) state of a recurrent nova cycle. The hot helium layer is still
		present.} \label{fig:TRho_unstable} \end{figure}

\begin{figure}
		\centering
		\includegraphics[width=\columnwidth]{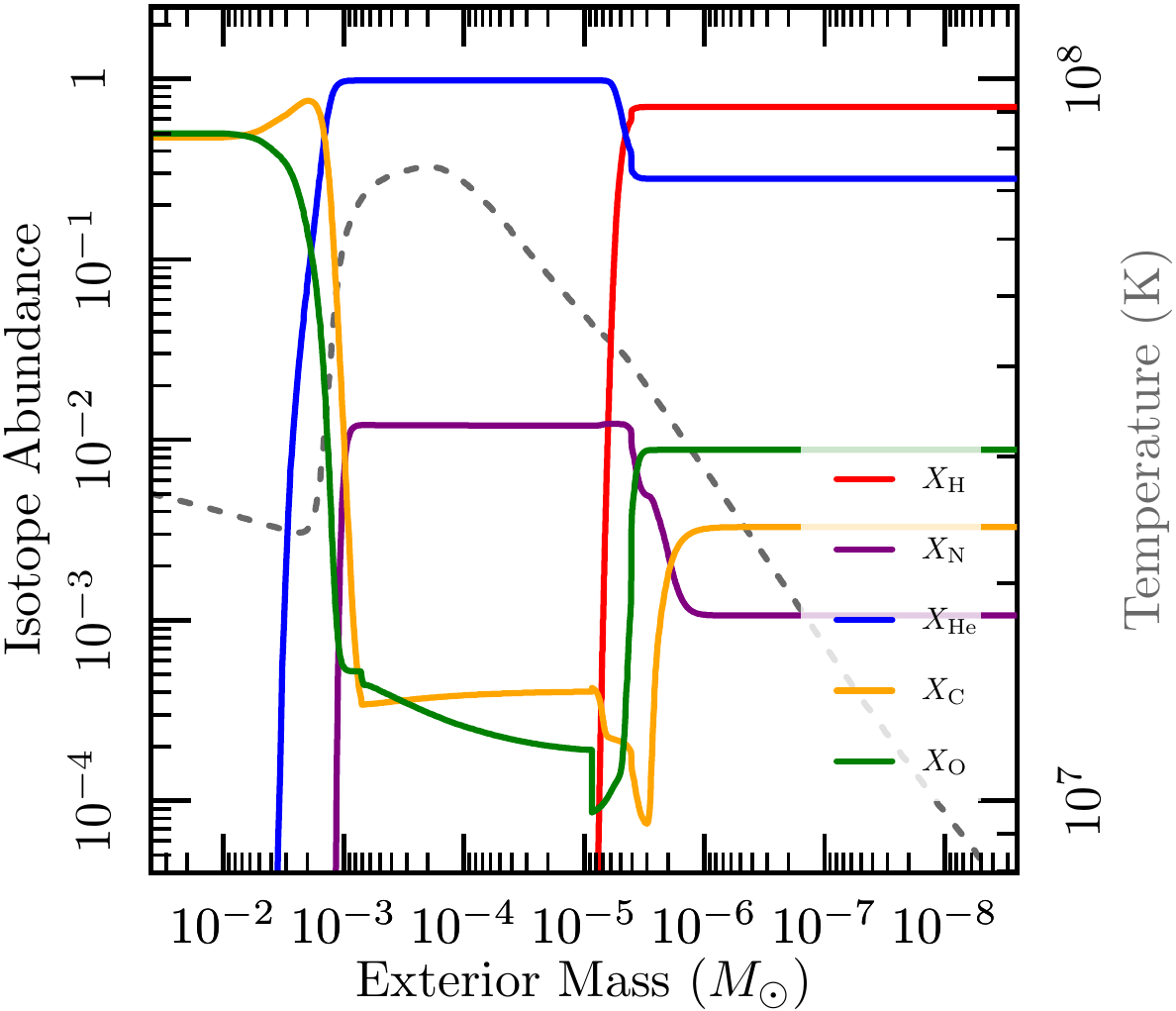}
		\caption{The abundance profile of a $1.00\ M_\odot$ WD accreting at
		\smash{$\dot{M}=1.7\times 10^{-7}\ M_\odot\,\mathrm{yr}^{-1}$}. The
		temperature profile has also been included to show that the hottest region
		is the layer of helium ash that dominates the burning conditions rather than
		the cooler core. Again, this profile is from the quiescent (pre-nova)
		state.}
		\label{fig:abundances_unstable}
	\end{figure}

	\cite{2004ApJ...600..390T} examined how classical novae (CNe) ignition masses
depend on \smash{$\dot{M}$}. Their analysis assumed that the core had reached
an equilibrium temperature due to prolonged thermal contact with the cycling
outer layers. We see that for sufficiently high \smash{$\dot{M}$}'s, the helium
layer retains a significant fraction of the thermal energy generated in a nova
event. So, for these \smash{$\dot{M}$}'s, the ignition mass and thus nearly all
other characteristics of a nova are independent of $T_c$. This trend is also
seen in the models with highest \smash{$\dot{M}$}'s in
\cite{2005ApJ...623..398Y}. The hot helium layer in the unstable models is
evident in Figures \ref{fig:TRho_unstable} and \ref{fig:abundances_unstable}
even while in the quiescent state. For CNe, CNO enrichment is seen in ejecta,
indicating that any helium layer from previous outbursts is mixed with the
hydrogen during the TNR and ejected along with a portion of the WD core. Thus,
we only expect the helium layer to be relevant at \smash{$\dot{M}$}'s near
\smash{$\dot{M}_{\mathrm{stable}}$} where mixing may cause helium dredge-up,
but not necessarily the removal of the entire layer. This would allow for the
gradual build-up of an insulating helium layer.
	
\begin{figure}
		\centering
		\includegraphics[width=\columnwidth]{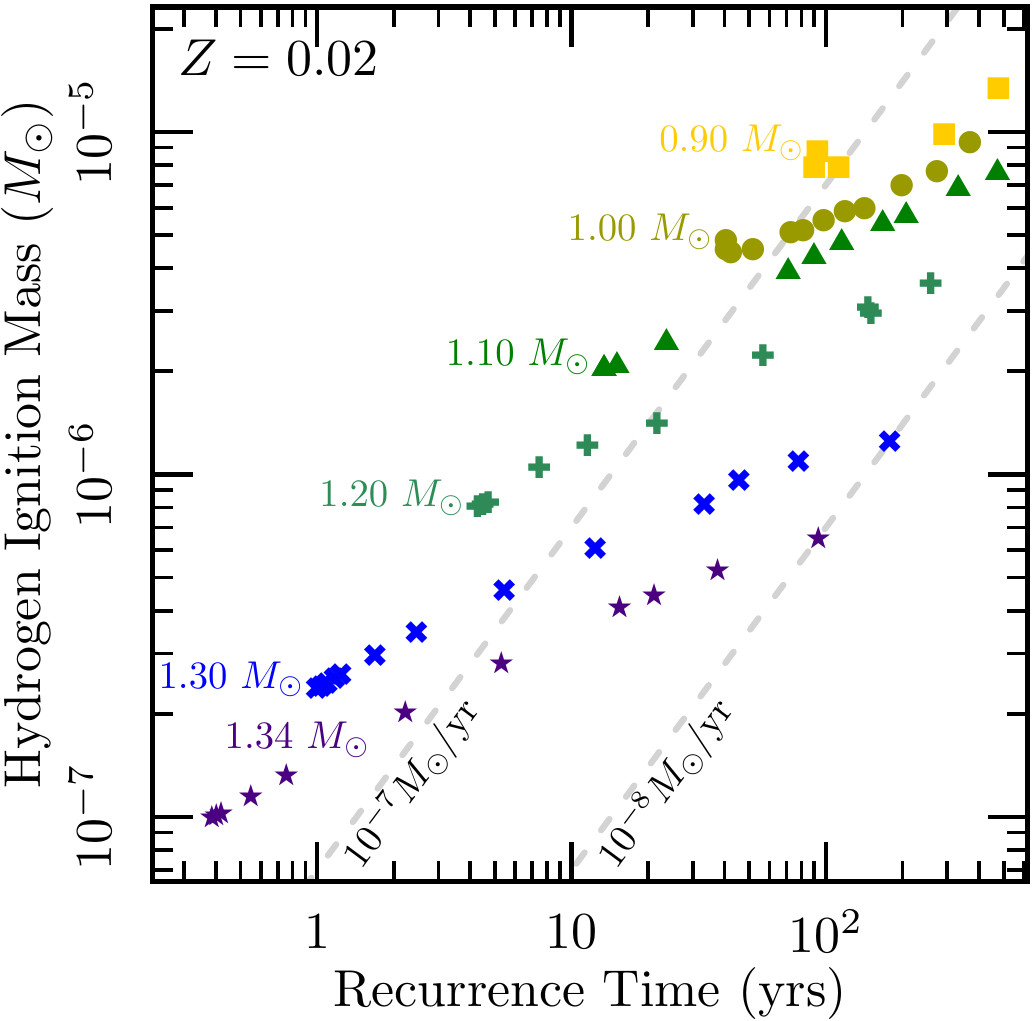}
		\caption{Ignition masses for high-mass WDs as a function of recurrence
		time. Lines of constant accretion rate (dashed) are also shown. The ignition
		masses reported here are the total amount of hydrogen accreted between
		outbursts, \smash{$X\dot{M} t_{\mathrm{recur}}$}, and thus necessarily lower
		than the total hydrogen mass present at the time of the nova eruption. We
		find that over the range of $M_{\mathrm{WD}}$ and \smash{$\dot{M}$}
		presented here, $\Delta M_{\mathrm{H,tot}}/\Delta M_{\mathrm{H,acc}}\approx
		1.2$.}
		\label{fig:ignition_masses}
	\end{figure}

	As \smash{$\dot{M}$} decreases, more time is allowed for the helium layer to
cool. This, in turn, causes the ignition mass to increase, since a higher
pressure is required to start a thermonuclear runaway at a lower temperature.
The trends for ignition masses in high-mass WDs are shown in
Figure~\ref{fig:ignition_masses}. Compared to their steadily burning
counterparts, the first unstable model (the left-most point for each mass in
Figure~\ref{fig:ignition_masses}) has a hydrogen ignition mass that is two to
three times larger than the stable hydrogen mass. Thus, the static hydrogen
masses from the steadily burning masses cannot be extrapolated into the RNe
regime to obtain recurrence times.
		

	\begin{deluxetable}{ccccccc}
		\tablecolumns{7}
		\tablewidth{0pc}
		\tablecaption{Properties of Recurrent Novae Just Below the Stability
			 						Boundary}
		\tablehead{
			\colhead{$M_{\mathrm{WD}}$}  & \colhead{$\dot{M}$} &
				\colhead{$t_{\mathrm{recur}}$} &
				\colhead{$t_{\mathrm{sweep}}$\tablenotemark{a}} &
				\colhead{$T_{\mathrm{He}}$\tablenotemark{b}} & 
				\colhead{$T_{\mathrm{peak}}$\tablenotemark{c}}\\
			\colhead{($M_\odot$)} & \colhead{($10^{-7}M_\odot\,\mathrm{yr}^{-1}$)} &
				\colhead{(yr)} & \colhead{(yr)} & \colhead{($10^7\,\mathrm{K}$)} &
				\colhead{($10^7\,\mathrm{K}$)}
		}
		\startdata
		0.51 & 0.24 & 4200 & 2810 & 5.0 & 8.3\\
		0.60 & 0.46 & 1300 & 720 & 5.8 & 8.8\\
		0.70 & 0.70 & 480 & 226 & 6.4 & 9.3\\
		0.80 & 1.00 & 200 & 88.7 & 7.2 & 10.0\\
		0.90 & 1.25 & 90 & 35.4 & 7.0 & 10.7\\
		1.00 & 1.7 & 40 & 15.6 & 8.0 & 11.1\\
		1.10 & 2.34 & 13.4 & 6.08 & 8.0 & 12.0\\
		1.20 & 2.7 & 4.3 & 2.24 & 9.9 & 13.1\\
		1.30 & 3.4 & 1.00 & 0.534 & 9.1 & 14.3\\
		1.34 & 3.7 & 0.39 & 0.226 & 10.0 & 15.1
		\enddata
		\label{tab:2}
		\tablenotetext{a}{Time for the stable model of the same $M_{\mathrm{WD}}$
		to burn through one full layer of hydrogen.}
		\tablenotetext{b}{Peak temperature during the low-luminosity state in the
		helium layer.}
		\tablenotetext{c}{Peak temperature during the outburst event.}		
	\end{deluxetable}

\section{Comparisons to Other Studies} 
\label{sec:comparisons_to_other_studies}
\begin{figure}[ht]
		\centering
		\includegraphics[width=\columnwidth]{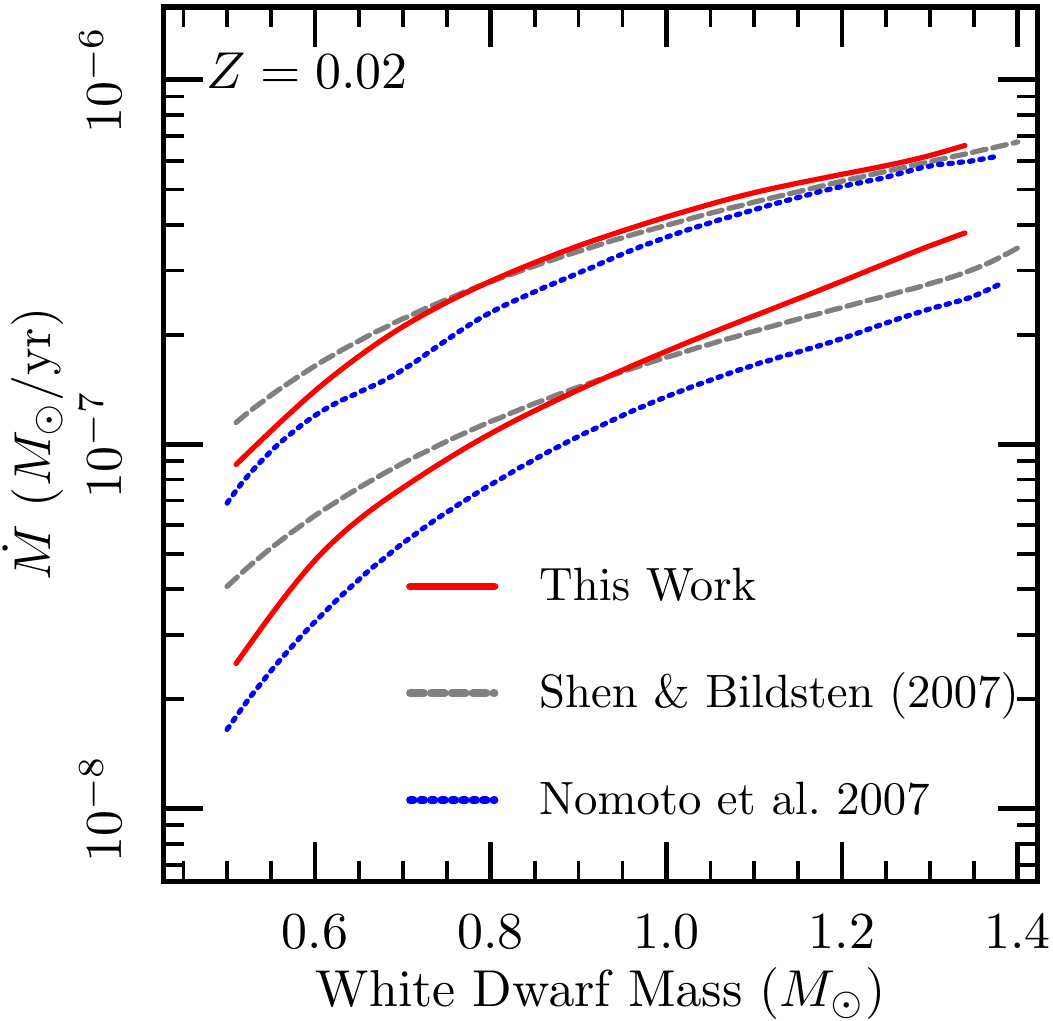}
		\caption{Stability regimes of this work (red solid line),
		\cite{2007ApJ...663.1269N} (dotted blue line), and
		\cite{2007ApJ...660.1444S} (dashed gray line).}
		\label{fig:mMdot}
	\end{figure}

	\cite{1980A&A....85..295S}, \cite{2007ApJ...660.1444S}, and
\cite{2007ApJ...663.1269N} all used linear stability analysis to test the
stability of constructed steady-state burning models. This is a
time-independent method that serves as a complementary check to our
time-dependent calculations. We find that our stability region shown in
Figure~\ref{fig:2} is largely consistent with these results and plot our
results along with those of \cite{2007ApJ...663.1269N} and
\cite{2007ApJ...660.1444S}. Each of these is plotted in Figure~\ref{fig:mMdot},
demonstrating the agreement between various techniques.
	
	\cite{2007ApJ...660.1444S} studied a one-zone model for stability at
various accretion rates and compared their results favorably to those of
\cite{2007ApJ...663.1269N}, noting that the discrepancy at lower masses was
likely due to their assumption of the burning layer being the mass within a
scale height. \cite{2007ApJ...663.1269N} compute the entire stellar model, but
assumed a discontinuous transition from solar composition to nearly pure helium
(Nomoto 2012, private communication). We, however, observe a transition zone
where most of the burning is occurring, so accurate comparisons are not
possible. Nonetheless, Figure~\ref{fig:mMdot} demonstrates agreement in the
stability boundary between the linear stability analysis and time-dependent
calculations.
	
	 Other recent time-dependent studies of accreting WDs in and near the
	 stable-burning regime have been carried out by \cite{1982ApJ...259..244I},
	 \cite{1989ApJ...341..299L}, \cite{1998ApJ...496..376C}, and
	 \cite{2005ApJ...623..398Y}. The thorough analysis in
	 \cite{1982ApJ...259..244I} shows that recurrence times change over the course
	 of several flashes. Hence, we only compare to simulations that computed
	 through multiple flashes to mitigate the effect of initial
	 condition choices. We now compare our simulations wherever possible.
	 
\cite{1982ApJ...259..244I} studied a 1 $M_\odot$ WD accreting in a quasi-static
(hydrostatic) approximation for \smash{$\dot{M}$}'s near and in the
stable-burning regime. He assumed $X=0.64$ in the accreted material and also
neglected mass loss. Both of these assumptions should lead to longer recurrence
times. The lowered hydrogen composition lowers the CNO energy generation rate,
requiring a higher pressure/temperature to get to the same level of burning as
would be expected if $X=0.70$. The lack of mass loss greatly affects the time
the WD spends at high luminosities, since it must burn through most of the
accreted envelope rather than removing most of it through winds. At high
\smash{$\dot{M}$}'s, the time spent on the high-luminosity branch is comparable
to the time spent in quiescence, so ignoring mass loss will lead to appreciably
longer recurence times. Additionally, the cores for \cite{1982ApJ...259..244I}
were typically much hotter than ours, exceeding the temperature of the
quiescent helium layer from our models. We expect this would act to decrease
the recurrence time since lower pressures (and thus accreted masses) are
required at higher temperatures to trigger a TNR. Finally,
\cite{1982ApJ...259..244I} must certainly have used different opacities, which
would affect the structure of the accreted envelope.

	At first, data for a stripped AGB core with $T_c\approx 3\times
10^8\,\mathrm{K}$ is presented. For $\dot{M}=2.5\times
10^{-7}\,M_\odot\,\mathrm{yr}^{-1}$, he reports steady and stable burning,
which we also observe. At \smash{$\dot{M}=1.5\times
10^{-7}\,M_\odot\,\mathrm{yr}^{-1}$}, he observes recurrence times at around 72
years, though they are evidently still increasing in his Figure 6. In contrast,
our corresponding model had a recurrence time of 42 years. His model has
approximately 36 years of intense hydrogen burning, whereas ours burns for only
17, indicating that mass loss is responsible for the much of the discrepancy.
He also displays data for the same hot WD as well as one whose core has gone
into a ``steady-state'' temperature configuration ($T_c\approx 7\times
10^7\ \mathrm{K}$) accreting at \smash{$\dot{M}=1.5\times 10^{-8}\, M_\odot\,
\mathrm{yr}^{-1}$}. These exhibit recurrence times of $\approx 550-650$ years,
though it is apparent in his Figures 7 and 11 that the recurrence times have
not grown to their asymptotic values yet. We observe longer recurrence times at
$\approx 1400$ years. In this case, the neglect of mass loss is likely
unimportant since the accretion phase is much longer than the outburst phase,
but the higher central temperatures of both WDs are pushing the recurrence
times down relative to ours.
	
	\cite{1989ApJ...341..299L} simulated a $1.0\ M_\odot$ WD with $T_c=10^8\
\mathrm{K}$ accreting at $10^{-8}$, $10^{-7}$, and $10^{-6}\
M_\odot\,\mathrm{yr}^{-1}$. Again, this core temperature is even hotter than
the stable burning temperature of the steady burners, so it will influence
ignition masses. Thus, we would expect their results to exhibit shorter
recurrence times and a lower stability boundary. Additionally, the accreted
material in their simulations had $X=0.7$ and $Z=0.03$. Their M8 model, at
\smash{$\dot{M} = 10^{-8}\ M_\odot\,\mathrm{yr}^{-1}$} exhibited repeated
hydrogen flashes with $t_{\mathrm{recur}} = 1520$ years and an ignition mass
(total mass present above the helium layer) of $\Delta
M_{\mathrm{ign}}=1.7\times 10^{-5}\ M_\odot$. In our corresponding model which
had the lower core temperature but the same \smash{$\dot{M}$}, we find
$t_{\mathrm{recur}} = 2220$ years and an ignition mass of $\Delta
M_{\mathrm{ign}}=2.6\times 10^{-5}\ M_\odot$, which given the core temperature
for such a relatively low \smash{$\dot{M}$}, is a plausible difference. For
their M7 model, which accreted at $10^{-7}\ M_\odot\,\mathrm{yr}^{-1}$, they
found $t_{\mathrm{recur}} = 135$ years, whereas our corresponding model gives
$t_{\mathrm{recur}} = 73$ years. Their M6 model, accreting at $10^{-6}\
M_\odot\,\mathrm{yr}^{-1}$ expanded to red giant proportions, as did ours.
	
	\cite{1998ApJ...496..376C} studied WDs with $M_{\mathrm{WD}} = 0.516\
M_\odot$ and $M_{\mathrm{WD}} = 0.80\ M_\odot$ accreting at rates comparable to
the stable burning regime. Their accreting matter has $X=0.7$ and $Z=0.02$, and
the core temperatures are well below their observed helium layer temperatures,
so we expect their models to compare more favorably to ours. It appears that
there was no mass loss prescription applied by \cite{1998ApJ...496..376C}. We
don't anticipate this causing any significant differences with our results
since the stable burning lifetimes for the configurations in question are small
compared to the accretion timescales. Additionally, the opacities used in
\cite{1998ApJ...496..376C} are taken from older Los Alamos tables that they
claim are very similar to the OPAL opacities. They simulated a 0.516 $M_\odot$
WD accreting at rates of $2\times 10^{-8}$, $4\times 10^{-8}$, $6\times
10^{-8}$, $10^{-7}$, and $10^{-6}\ M_\odot\,\mathrm{yr}^{-1}$. They observe
that the model with \smash{$\dot{M}=2\times 10^{-8}\
M_\odot\,\mathrm{yr}^{-1}$} exhibits hydrogen flashes with $t_{\mathrm{recur}}
= 5400$ years, whereas our $0.51\ M_\odot$ model at the same \smash{$\dot{M}$}
has $t_{\mathrm{recur}} = 5040$ years. For \smash{$\dot{M}=4\times 10^{-8}$},
$6\times 10^{-8}$, and $1\times 10^{-7}\ M_\odot\,\mathrm{yr}^{-1}$, they
observe steady burning, which is mostly consistent with our results, though we
found that for our $0.51\ M_\odot$ WD, an \smash{$\dot{M}$} of $10^{-7}\
M_\odot\,\mathrm{yr}^{-1}$ resulted in a red giant configuration. Their WD is
slightly more massive, and given that we find that \smash{$\dot{M}=9\times
10^{-8}\ M_\odot\,\mathrm{yr}^{-1}$} gives stable burning on our $0.51\
M_\odot$ WD, the discrepancy seems plausible. Finally, both
\cite{1998ApJ...496..376C} and we observe a red giant phase for
\smash{$\dot{M}=1\times 10^{-6}\ M_\odot\,\mathrm{yr}^{-1}$}.
	
		For their $0.80\ M_\odot$ WD, \cite{1998ApJ...496..376C} ran simulations
		with \smash{$\dot{M}=10^{-8}$}, $4\times 10^{-8}$, $10^{-7}$, $1.6\times
		10^{-7}$, and $4\times 10^{-7}\ M_\odot\,\mathrm{yr}^{-1}$. For the lowest
		three \smash{$\dot{M}$}'s, they found $t_{\mathrm{recur}} = $ 3110 years,
		483 years, and 204 years, respectively. For the same mass and
		\smash{$\dot{M}$}'s, we find $t_{\mathrm{recur}} = $ 3370, 596, and 200
		years, respectively. We both observe steady burning at
		\smash{$\dot{M}=1.6\times 10^{-7}\ M_\odot\,\mathrm{yr}^{-1}$}, and we both
		observe a red giant configuration at \smash{$\dot{M}=4\times 10^{-7}\
		M_\odot\,\mathrm{yr}^{-1}$}.
	
	Finally, we compare to \cite{2005ApJ...623..398Y}, who simulated accreting
WDs with masses of 0.65, 1.00, and 1.25 $M_\odot$ (among others that we do not
compare to). For each of these masses, they accreted matter at rates of
$10^{-8}$ and $10^{-7}\ M_\odot\,\mathrm{yr}^{-1}$ (again with many more at
lower \smash{$\dot{M}$}'s that aren't applicable to our study). They also
varied the core temperature between $1\times 10^7\ \mathrm{K}$ and $5\times
10^{7}\ \mathrm{K}$. It shouldn't affect our results, but we will compare only
with the $T_c=3\times 10^7\ \mathrm{K}$ results. Finally, they employed an
optically thick, supersonic wind as a mass loss prescription
\citep{1995ApJ...445..789P} and allowed for convective overshoot, as
evidenced by their metal-enriched ejecta.
	
	At no point is stable burning reported in \cite{2005ApJ...623..398Y}, though
it seems that simulations with no mass loss correspond to our red giant or
stable configurations (a period is still reported in their Table 3). For
$M_{\mathrm{WD}}=0.65\ M_\odot$, $t_{\mathrm{recur}} =$ 10200 and 254 years are
reported for \smash{$\dot{M} = 10^{-8}$} and $10^{-7}\
M_\odot\,\mathrm{yr}^{-1}$, respectively. Using our own $0.65\ M_\odot$ WD, we
find a recurrence time of 7800 years for \smash{$\dot{M} = 10^{-8}\
M_\odot\,\mathrm{yr}^{-1}$} and stable burning for \smash{$\dot{M} = 10^{-7}\
M_\odot\,\mathrm{yr}^{-1}$}. For the 1.00 $M_\odot$ case, the two reported
$t_{\mathrm{recur}}$'s are 2030 and 87.4 years, whereas ours are
$t_{\mathrm{recur}}=$ 2216 and 72.6 years. Finally, their 1.25 $M_\odot$ WDs
give $t_{\mathrm{recur}}=$ 384 and 19.6 years. Our 1.25 $M_\odot$ WD models
indicate $t_{\mathrm{recur}}=$ 258 and 14.4 years at these \smash{$\dot{M}$}'s.
For the higher \smash{$\dot{M}$}'s, there is little to no metal enrichmeent in
the ejecta and only minor helium enrichment, so we expect the reasonable
agreement in most of the calculations. The exception is the $M=1.25\,M_\odot$,
\smash{$\dot{M}=10^{-8}\,M_\odot\,\mathrm{yr}^{-1}$} calculation, where the
ejecta in \cite{2005ApJ...623..398Y} is significantly metal-enriched,
indicating dredge-up from the core. It's not immediately obvious why our
calculation with no enrichment has a shorter recurrence time, since CNO burning
should start more easily with an enriched base layer.
  

\section{Post-Outburst Novae} 
\label{sec:post_outburst_novae}
In addition to the models computed for stability analysis, we also ran models
with $M_{\mathrm{WD}}=0.6\,M_\odot$, $1.0\,M_\odot$, $1.1\,M_\odot$,
$1.2\,M_\odot$, $1.3\,M_\odot$, and $1.34\,M_\odot$ at a lower accretion rate of
\smash{$\dot{M}=10^{-9}\,M_\odot\,\mathrm{yr}^{-1}$} to study the stable
burning phase after a classical nova (CN). For mass loss, we used both the
super Eddington wind prescription described earlier as well as Roche lobe
overflow (RLOF) by putting the WDs in a binary systems with Roche lobe radii
between $R_{\mathrm{RL}}=0.4\,R_\odot$ and $1.0\ R_\odot$. Our choice for the
accretion rate, masses, and orbital separation was motivated by the study of
the classical novae population by \cite{2005ApJ...628..395T}. They showed that
the observed orbital period distribution of the CNe was consistent with
expectations of the mass transfer rate history of cataclysmic variables. This
implied that the most often observed CNe would be those in 4-7 hours orbital
periods with a mass transfer rate driven by magnetic braking at
$10^{-9}\,M_\odot\,\mathrm{yr}^{-1}$. These tight orbits then enable Roche lobe
overflow when the WD undergoing the CN reaches a photospheric radius
$R_{\mathrm{WD}}= R_{\mathrm{RL}}$, triggering the mass loss from the WD that
creates a common envelope. Within \texttt{MESA} this mass loss is simulated by
eliminating any mass beyond the Roche lobe radius, effectively demanding that
the WD photosphere not exceed $R_{\mathrm{RL}}$. It is simply the hydrostatic
expansion of the actively burning layer that pushes the outer layers beyond
$R_{\mathrm{RL}}$. Once the hydrogen layer mass has reduced to a value where
$R_{\mathrm{WD}}\lesssim R_{\mathrm{RL}}$, the mass loss ends and the period of
prolonged stable burning ensues. Since the ignition masses are smaller on more
massive WDs, the expectation is that, even though rarer, more massive WDs will
be more prevalent in the observed population.

\cite{1994ApJ...437..802K} have accounted for mass loss in novae through
optically thick winds driven by an opacity bump at $\log T(\mathrm{K})\approx
5.2$ from the OPAL tables. This bump in opacity should cause a decrease in the
Eddington luminosities, making our super Eddington wind prescription a
plausible mass loss mechanism. We present results using both mass loss
mechanisms independently, but it's likely that some combination of winds and
Roche lobe overflow are present in actual novae. Finally, we again neglect
convective dredge-up and the accompanying metal enrichment of the burning layer.

\begin{figure}
	\centering
	\includegraphics[width=\columnwidth]{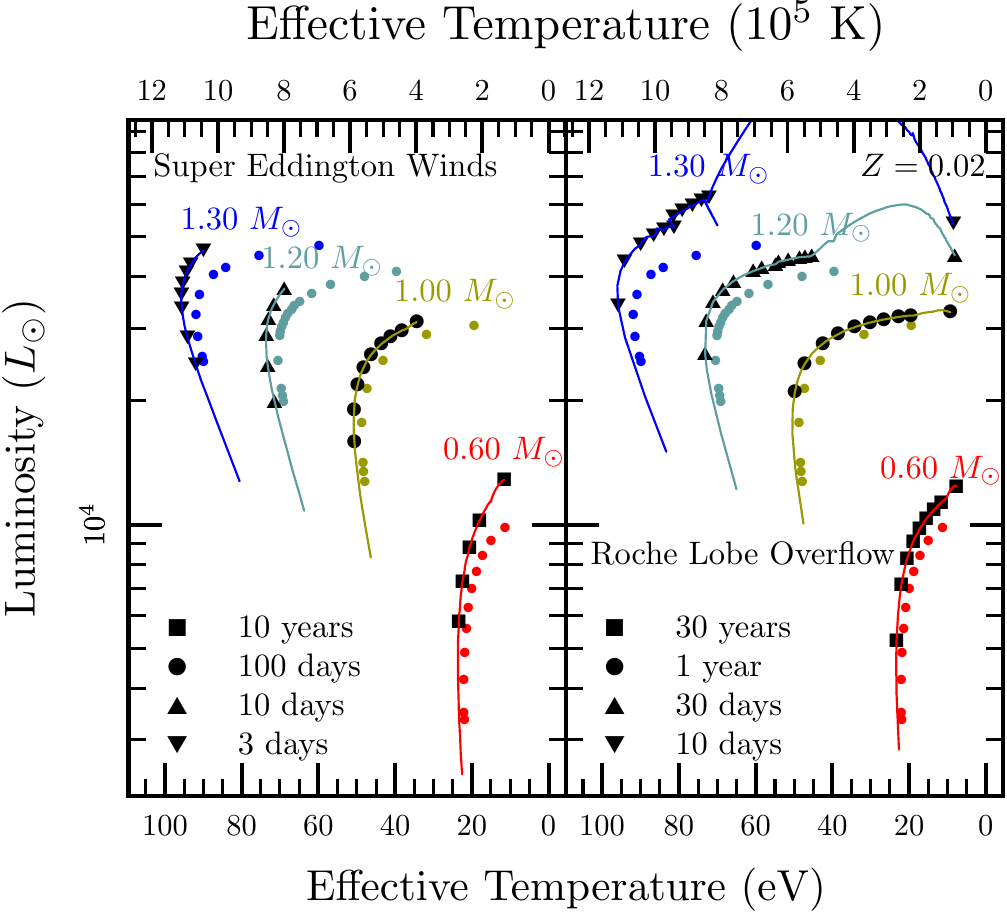}
	\caption{The paths of the CNe through the HR diagram after outburst (lines)
with the static positions of the steadily-burning models presented earlier
(circles). Models using super Eddington winds are on the left, and those using
RLOF are on the right. For both mass loss mechanisms, we divide
the tracks into equal time portions with markers whose time lengths are
indicated in the legend. }
	\label{fig:novae_hr}
\end{figure}

The evolutionary tracks of our $0.6\,M_\odot$, $1.0\,M_\odot$, $1.2\,M_\odot$,
and $1.3\,M_\odot$ models just after the end of mass loss are shown on the HR
diagram with respect to the stable burners of \S
\ref{sec:steadily_burning_models} for both mass loss prescriptions in
Figure~\ref{fig:novae_hr}. There we show the CNe as lines with markers on them
denoting equal time steps after mass loss has ended. For instance, the RLOF
0.60 $M_\odot$ model burns steadily for approximately 250 years, whereas our
RLOF 1.20 $M_\odot$ model only does so for approximately a year. The duration
of the supersoft source (SSS) phase is clearly dependent on both the WD mass
and the amount of mass ejected. The super Eddington wind models tend to start
``later'' in the steady-state locus since the super Eddington winds remove more
mass than RLOF. As a result, those turn-off times are always shorter than the
those of the corresponding RLOF models. We found there is some cutoff mass that
depends on orbital parameters below which novae fill their Roche lobe before
the luminosity goes super Eddington. For the models shown, the 0.6 $M_\odot$
and 1.0 $M_\odot$ fill their Roche lobes before going super Eddington, so the
RLOF prescription is likely more accurate. For the higher masses, the Roche
lobe is bigger, allowing for greater expansion, leading to lower envelope
temperatures, greater opacities, and thus a lower Eddington limit. The super
Eddington models for these higher masses never expanded to the Roche lobe
radius set for the RLOF models. Additionally, we can see that during the
contracting phase, the RLOF 1.2 $M_\odot$ and 1.3 $M_\odot$ WDs are super
Eddington (the upwards excursion), so the mass loss in those cases is certainly
a lower limit.

\begin{figure}
	\centering
	\includegraphics[width=\columnwidth]{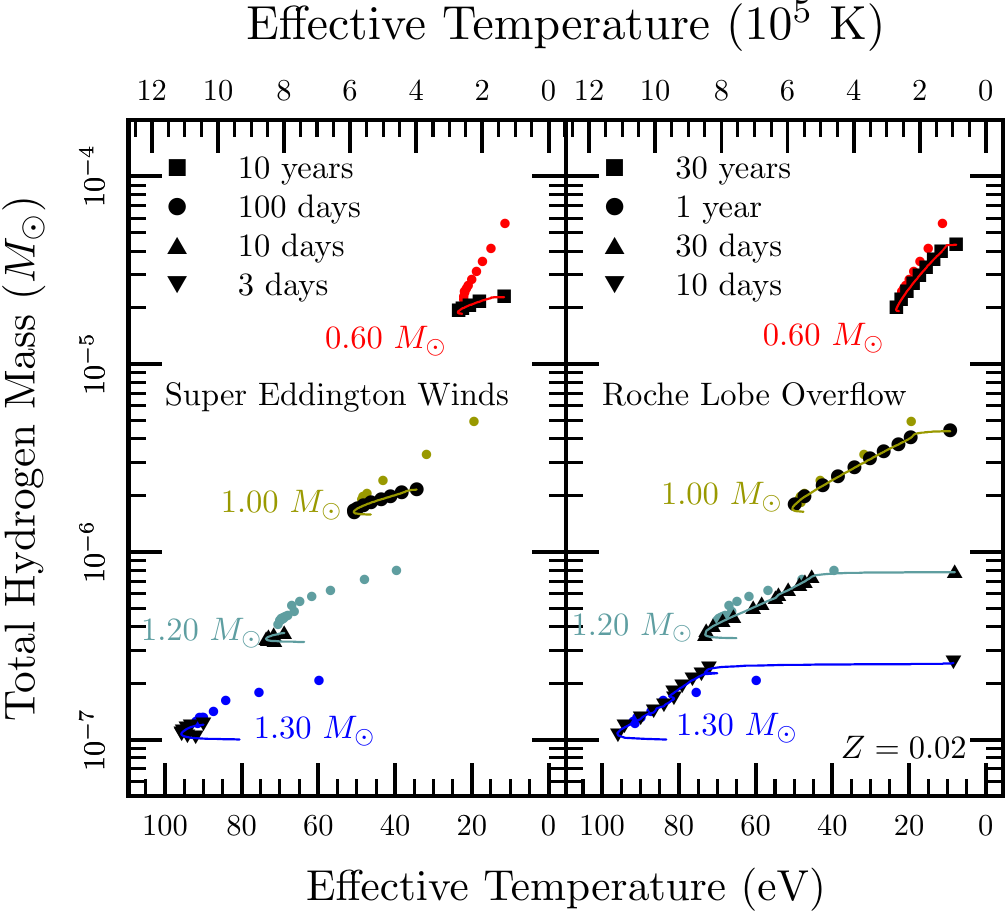}
	\caption{The hydrogen mass against the effective temperature of post-outburst
CNe (lines) compared to steady burners of the same mass (circles). Models using
super Eddington winds are on the left and those using RLOF are on the right.
The RLOF novae pass through phases closely resembling their steadily-burning
counterparts at the same effective temperatures, but the stronger mass loss
from super Eddington winds cause novae to start stable burning at a much lower
envelope mass than the corresponding RLOF models. Equal time markers are the
same as mentioned in Figure~\ref{fig:novae_hr}.}
	\label{fig:mh_kTeff_with_novae}
\end{figure}

After mass loss (on the far red end of each evolutionary track), each CN passes
near or directly through the locus of stably burning phases corresponding to
its mass. However, at an accretion rate of \smash{$\dot{M}=1.0\times
10^{-9}\,M_\odot\,\mathrm{yr}^{-1}$}, the stable burning consumes hydrogen
faster than it is accreted. Thus, a CN passes through phases with a
progressively smaller hydrogen layer, tracing a path to and around the knee
until the layer becomes thinner than that of the critically stable WD
configuration. \cite{2005A&A...439.1061S} modeled CNe in the post-outburst
phase as a series of stably-burning WDs and tracked their evolution for four
envelope compositions. Their Figure 1 gives HR diagram paths as well as the
evolution of the hydrogen-rich layer in each of their modeled CNe. Their
asymptotic luminosities and effective temperatures for the most metal-poor
configuration (ONe25, at $Z_{\mathrm{env}} = 0.25$) agree well with our stable
burners, though we find that the CNe themselves follow tracks that are
marginally brighter and hotter than the corresponding stable burners. The
depleting hydrogen layer is very apparent in
Figure~\ref{fig:mh_kTeff_with_novae}, where we see that the WDs using RLOF
realize states with hydrogen masses and $T_{\mathrm{eff}}$'s very close to the
corresponding steadily burning WDs. However, the WDs using super Eddington
winds typically removed more mass than the RLOF models, so they ``skip'' some
or most of the steady-state configurations and instead start with a much lower
envelope mass. This disparity in the amount of fuel between the two
configurations at the same mass explains why the turn-off times are much
shorter for super Eddington winds than RLOF. In either cases, hydrogen burning
becomes an insignificant source of luminosity past the lowest \smash{$\dot{M}$}
stable burner state, and the WD then proceeds down the WD cooling track at
nearly constant radius and hydrogen mass. Comparing to the ONe25 model in
\cite{2005A&A...439.1061S}, we observe turn-off times that are always longer in
the RLOF and low-mass super Eddington cases. For the higher mass super
Eddington models, we observe marginally shorter turn-off times, likely driven
by the skipped steady-state modes. The overall trend is that nearly all
turn-off times in \cite{2005A&A...439.1061S} are shorter than ours due to the
significant metal enrichment of their envelopes, which is an important
difference we elaborate on later.

\begin{figure}
	\centering
	\includegraphics[width=\columnwidth]{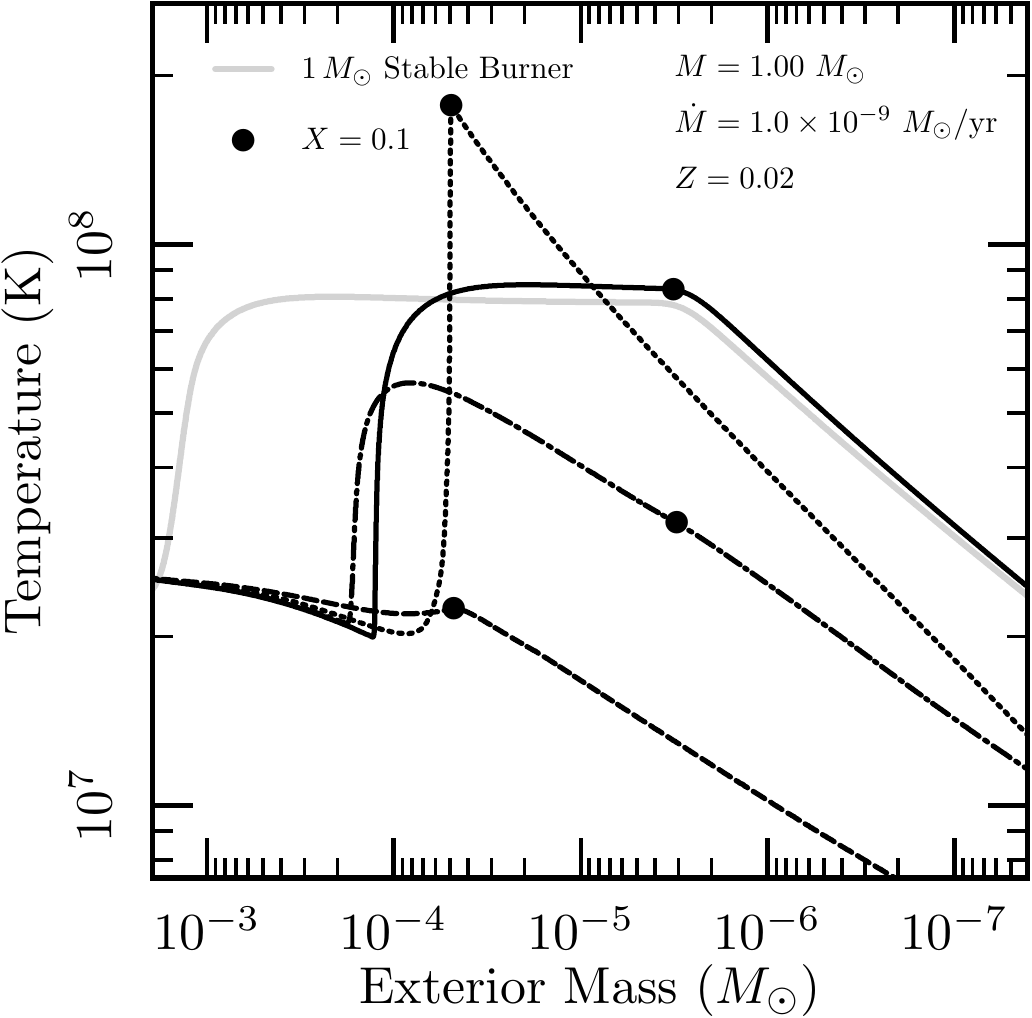}
	\caption{Time series of temperature profiles in a $1.00\ M_\odot$ WD
accreting solar material at \smash{$\dot{M}=1.0\times 10^{-9}\
M_\odot\,\mathrm{yr}^{-1}$} with RLOF for mass loss. The point shows the base
of the H layer. The long-dashed line is just prior to the outburst, when the
luminosity is low and due primarily to compressional heating. The dotted line
is the profile at the time of peak hydrogen burning, with a vigorously
convective layer extending from the point of peak burning. The solid line is
taken from the time at which $T_{\mathrm{eff}}$ is at a maximum, marking the
end of the stable burning phase. The dash-dotted line is from shortly after the
stable burning ceases, as the envelope is cooling off and accumulating
hydrogen. The gray line is the profile of a $1.00\ M_\odot$ WD accreting at the
lower stable limit.}
	\label{fig:time_series_profiles}
\end{figure}

Figure~\ref{fig:time_series_profiles} shows the temperature profile of the
$1.0\ M_\odot$ CN at four distinct stages: the low luminosity accreting state,
the peak of hydrogen burning during the TNR, the point of highest
$T_{\mathrm{eff}}$, which is near the end of stable burning, and the
cooling/accumulating phase just after stable burning has ceased. For each
profile, the location of the base of the hydrogen-burning layer (here
approximated as the location where $X=0.1$) is marked. For comparison, the
lowest-\smash{$\dot{M}$} $1.00\ M_\odot$ stable burner is also shown in the
gray line. As the hydrogen accumulates in the low-luminosity state, the profile
is somewhat similar to a cooling WD, albeit with some heat still left over in
the helium layer as well as some energy generation due to the compressional
losses from accretion (see Figures 26 and 27 in \cite{2013ApJS..208....4P}).
Once the pressure at the base of the hydrogen reaches a critical threshold, the
thermonuclear runaway (TNR) ensues, raising the temperature at the base to
almost $2\times 10^8\ \mathrm{K}$, which in turn drives a convective zone in
the hydrogen layer. The radius then expands, triggering Roche lobe overflow
until the envelope's thermal structure is reorganized so that it can carry the
luminosity from the hydrogen burning. It then enters the stable burning phase,
during which we see a temperature profile in the hydrogen-rich layer that is
very similar to a steadily burning WD. Note that between the TNR and the SSS
phase, approximately 90 percent of the hydrogen layer has been lost. A small
portion of this is due to the stable burning, but the majority is due to RLOF.
After stable burning ceases, the envelope cools and accretes hydrogen until the
next TNR repeats the process.

\begin{figure}
	\centering
	\includegraphics[width=\columnwidth]{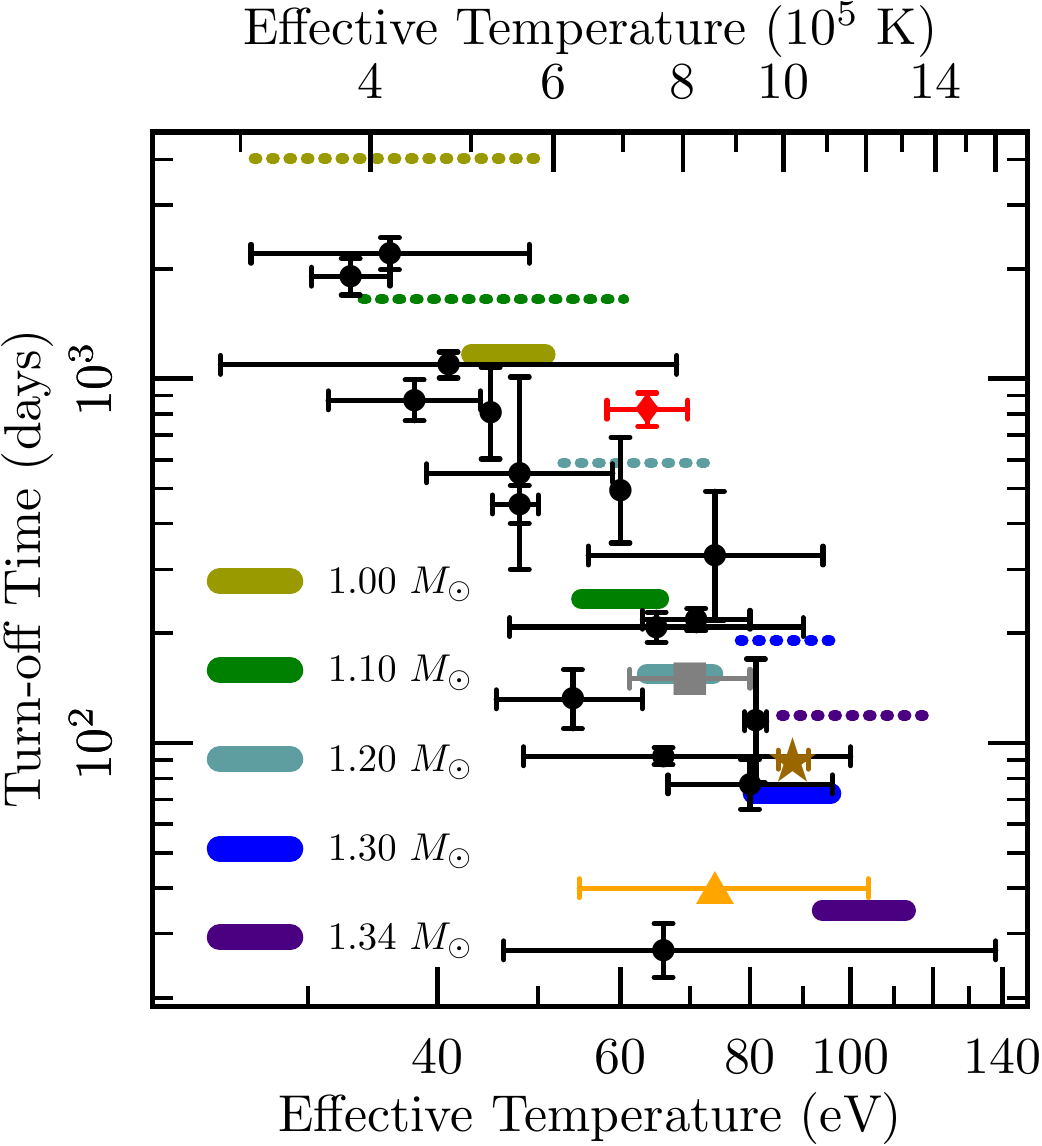}
	\caption{Turn-off time against $kT_{\mathrm{eff}}$ for observed CNe from the
catalogue of M31 CNe in \cite{Henze:2011bn} (black dots), the CN in globular
cluster Bol 126 in M31 \citep{2013A&A...549A.120H} (orange triangle), V4743 Sgr
\citep{2010ApJ...717..363R} (red diamond), the recurrent nova RS Oph
\citep{2011ApJ...727..124O} (brown star), HV Ceti \citep{2012A&A...545A.116B}
(gray square), as well as the results of our work. Solid lines represent the
super Eddington wind models and the dotted lines represent RLOF models. The
range of effective temperatures shown for the computational models are the
temperatures during the latter 70\% of the stable burning period. Note that the
$T_{\mathrm{eff}}$ for the \texttt{MESA} simulations are from the
Stefan-Boltzmann law given a luminosity and a photospheric radius.
$T_{\mathrm{eff}}$'s from \cite{Henze:2011bn,2013A&A...549A.120H} are blackbody
approximations taken from X-Ray spectra, and $T_{\mathrm{eff}}$'s from the
galactic novae are from NLTE models of hot WD atmospheres. These temperatures
can differ by $\approx 10\%$ due to radiative transfer effects. }
	\label{fig:henze}
\end{figure}

Combined measurements of $T_{\mathrm{eff}}$ and the turn-off time of a CN
\citep{Henze:2011bn} can be used to infer the WD mass. Figure~\ref{fig:henze}
shows observed turn-off times and $T_{\mathrm{eff}}$'s in post-ouburst novae.
Included in Figure~\ref{fig:henze} are data from M31
\citep{Henze:2011bn,2013A&A...549A.120H} and galactic sources
\citep{2010ApJ...717..363R,2011ApJ...727..124O,2012A&A...545A.116B}.
For the \cite{Henze:2011bn} dataset, we've only included data that had reported
uncertainties rather than limit points, which discriminates against
longer-lived SSS phases since they are less likely to be observed from
beginning to end. We also plot our calculations, where we define the
turn-off time as the time between the beginning of mass loss and when the
luminosity falls below one quarter of the peak luminosity of the stable burners
of the same mass. This ending criteria isn't very crucial since
the luminosity evolution after the stable burning phase is very rapid compared
to time spent doing stable burning. For the computational models, we cannot
report a single $T_{\mathrm{eff}}$ since it increases through most of the SSS
phase. The turn-off time is well defined, so we report our results as a
horizontal line in Figure~\ref{fig:henze} with the effective temperatures being
those during the latter 70\% of stable burning (the SSS is likely obscured at
earlier times by the expanding ejecta shell).

The temperatures reported for the observed CNe are obtained either by
approximating an X-ray spectrum as a blackbody
\citep{Henze:2011bn,2013A&A...549A.120H} or through more sophisticated NLTE
simulations \citep{2010ApJ...717..363R, 2011ApJ...727..124O,
2012A&A...545A.116B}. These two methods can yield different temperatures by
$\approx 10\%$ (see Figure 4 from \citealt{Henze:2011bn}), so there will
necessarily be disagreement between CNe analyzed by the two different methods.
Our models do not account for dredge-up and the subsequent metal enrichment of
the ejecta and stably-burning envelope. This could cause two effects. First,
the enriched TNR could burn more vigorously, driving stronger mass loss and
thus shortening the turn-off time. Secondly, the remnant envelope after mass
loss will be metal-enriched and will thus burn through the remaining hydrogen
more quickly than if the same mass were at solar abundance, as shown in
\cite{2005A&A...439.1061S}. Both of these factors indicate that our turn-off
times are longest limits for the given mass loss prescriptions. We do not,
however, expect metal enrichment to alter the effective temperature of the
outburst, so these can still be used to constrain WD masses. Finally, the
observed data shown in Figure~\ref{fig:henze} are deficient in low-temperature
($kT_{\mathrm{BB}} < 30\,\mathrm{eV}$) events.

Such events do exist, but the available measured turn-off times for
them are lower limits since they have not been observed for a long enough time
to detect both turn-on and turn-off. Additionally, observing such events is
difficult due to absorption by interstellar neutral hydrogen and the overall
weaker X-Ray flux. Finally, such low-mass systems may be more numerous, but
since their recurrence times are significantly longer than their higher-mass
counterparts, they are observed less often. In Figure~\ref{fig:henze} we only
plot those events from \cite{Henze:2011bn} that have established uncertainties
in both the blackbody temperature and the turn-off time. The agreement between
theory and observation in this region of parameter space is strong, implying
that most of the novae with SSS phases that have established turn-off times and
blackbody temperatures have $M\geq M_\odot$. As X-ray monitoring of M31
continues, more SSS's will turn off and stacked pointings allow for detection
of fainter SSS's. Thus we will soon be able to probe more reliably into the
lower-mass regime (Henze 2013, private communication).

\begin{figure}
	\centering
	\includegraphics[width = \columnwidth]{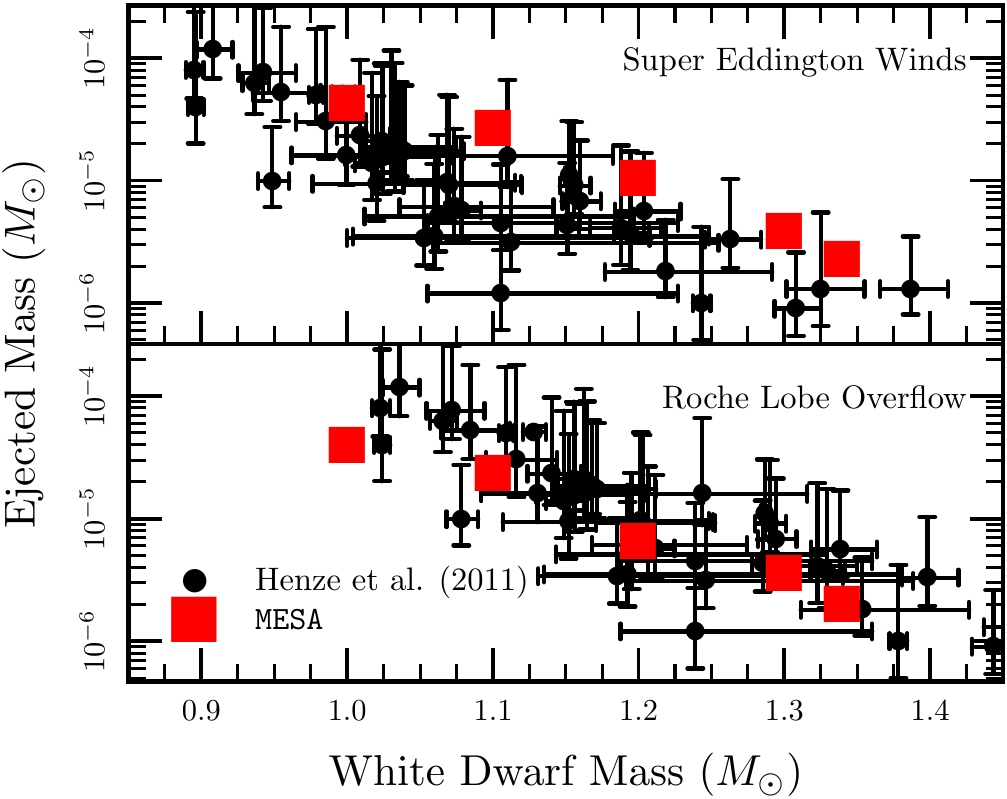}
	\caption{Ejected mass in a nova as a function of WD mass for both the
\texttt{MESA} simulations as well as the same catalogue of observations used in
Figure~\ref{fig:henze}. The masses for the observed data were obtained by using
a power law fit from the simulated data to convert turn-off times to masses.}
	\label{fig:m_ej}
\end{figure}

Due to the variability of $T_{\mathrm{eff}}$ during the SSS phase, it is not an
ideal tracer of WD mass on its own. The turn-off time, however, is a function
of the luminosity and hydrogen mass layer size, assuming the mass loss history
is known. Since we've seen that $\Delta M_{\mathrm{H}}$ decreases with
increasing $M_{\mathrm{WD}}$ while $L$ increases with increasing
$M_{\mathrm{WD}}$, the turn-off time should be a consistent tracer of WD mass
while also being relatively easy to measure. Using our high-mass CNe models,
including an additional $1.34\,M_\odot$ CN, we find power laws relating
turn-off time to WD mass given by $M_{\mathrm{WD}}=1.20\,M_\odot\,(513\
\mathrm{days}/t_{\mathrm{off}})^{0.081}$ (for RLOF) and $M_{\mathrm{WD}} =
1.20\,M_\odot\,(137\ \mathrm{days}/t_{\mathrm{off}})^{0.089}$ (for super
Eddington winds). We then apply this relation to the catalogue of
\cite{Henze:2011bn} to get corresponding WD masses to compare to the reported
ejection masses, which were inferred by \cite{Henze:2011bn} from the turn-on
time and the ejecta velocity. The results of this analysis are shown in
Figure~\ref{fig:m_ej}. We see the mapping from turn-off time to
$M_{\mathrm{WD}}$ gives a similar relation between WD mass and ejected mass as
the simulations for either mass loss prescription. Note though that the RLOF
law gives super-Chandrasekhar mass WDs for sufficiently low turn-off times,
which is a result of the under-prediction of mass loss in high-mass WDs in the
RLOF assumption.

\section{Concluding Remarks} 
\label{sec:concluding_remarks}
	We have presented stably burning WD models, found the lowest
\smash{$\dot{M}$}'s that permit such stable burning, and verified that they are
consistent with other time-dependent studies as well as time-independent linear
analysis studies of stability. We've shown that the hot helium ash left over
from hydrogen burning dominates the thermal structure of both stably and
unstably burning WDs at high \smash{$\dot{M}$}'s. This helium layer is
important because it sets the recurrence times for rapidly accreting recurrent
novae where dredge up is unable to reach the WD core, but it is also important
because it is likely to ignite unstably once it has grown large enough. The
mass of helium in the WD, $\Delta M_{\mathrm{He}}$, is not a static property of
a stably burning model. For the $1.00\ M_\odot$ example shown in
Figure~\ref{fig:abundances}, $\Delta M_{\mathrm{He}}\approx 1.5\times10^{-3}\
M_\odot$, but it will continue to grow at the accretion rate,
\smash{$\dot{M}$}, until the pressure and the temperature at the base become
high enough to initiate unstable helium burning. \cite{1980A&A....85..295S},
\cite{1989ApJ...342..430I}, and \cite{1998ApJ...496..376C} showed that the
stable burning regimes for hydrogen and helium are mutually exclusive for the
case of solar composition accretion. However, \cite{2004A&A...425..207Y} offer
a way to merge the two stability regimes if a large amount of differential
rotation is allowed in the burning shells. Our calculations assume no rotation,
and we observed unstable helium burning for WDs that were allowed to
continuously accrete.

	 Finally, we showed how CNe pass through the stably burning phases after
	 their outburst and subsequent mass loss. The duration of this SSS phase is
	 highly sensitive to the mass of the underlying WD, spanning for hundreds of
	 years for $M_{\mathrm{WD}}\approx 0.60\ M_\odot$ to mere tens of days for
	 $M_{\mathrm{WD}}\approx 1.30\ M_\odot$. This variety of durations indicates a
	 mapping from observed turn-off times to WD mass, though a study of the
	 effects of metal enhancement and a better understanding of mass loss is
	 necessary to get a more precise relationship.
	
	We thank Ken Nomoto for helpful discussions regarding the calculations in
\cite{2007ApJ...663.1269N}. Additionally we thank Pablo Marchant for his
useful binary Roche lobe overflow routines that were used in this work and
Jeno Sokoloski for consultation regarding RNe observations. Finally, we thank
Martin Henze and the referee for very helpful comments. Most of the simulations
for this work were made possible by the Triton Resource. The Triton Resource is
a high performance research computing system operated by the San Diego
Supercomputer Center at UC San Diego. This work was supported by the National
Science Foundation under grants PHY 11-25915, AST 11-09174 and AST 12-05574.

\bibliographystyle{apj}
\bibliography{stablerefs}
\end{document}